\documentclass[twocolumn]{aastex7}
\usepackage{amsmath}
\usepackage{booktabs}
\usepackage{multirow}

\newcommand\dd{\mathrm{d}}

\newcommand\mv{\sum m_\nu}
\newcommand\mveff{\sum m_{\nu, \rm eff}}

\newcommand\neff{N_\mathrm{eff}}

\newcommand\wcdm{w_0w_a\mathrm{CDM}}

\newcommand{\fx}{\textcolor{black}}

\shorttitle{J-PAS and PFS forecasts}
\shortauthors{Qin, Wang \& Zhao et al.}
\submitjournal{ApJ}

\graphicspath{{./}{fig/}}

\begin{document}

\title{J-PAS and PFS surveys in the era of dark energy and neutrino mass measurements}

\author{Fuxing Qin}
\affiliation{National Astronomical Observatories, Chinese Academy of Sciences, Beijing, 100101, P.R.China}
\affiliation{School of Astronomy and Space Sciences, University of Chinese Academy of Sciences, Beijing, 100049, P.R.China}
\email[]{qinfx@bao.ac.cn}

\author[0000-0001-7756-8479]{Yuting Wang}
\affiliation{National Astronomical Observatories, Chinese Academy of Sciences, Beijing, 100101, P.R.China}
\affiliation{Institute for Frontiers in Astronomy and Astrophysics, Beijing Normal University, Beijing 102206, China}
\email[show]{Yuting Wang: ytwang@nao.cas.cn}

\author[0000-0003-4726-6714]{Gong-Bo Zhao}
\affiliation{National Astronomical Observatories, Chinese Academy of Sciences, Beijing, 100101, P.R.China}
\affiliation{School of Astronomy and Space Sciences, University of Chinese Academy of Sciences, Beijing, 100049, P.R.China}
\affiliation{Institute for Frontiers in Astronomy and Astrophysics, Beijing Normal University, Beijing 102206, China}
\email[show]{Gong-Bo Zhao: gbzhao@nao.cas.cn}

\author{Antonio J. Cuesta}
\affiliation{Departamento de Física, Universidad de Córdoba, Campus Universitario de Rabanales, Ctra. N-IV Km. 396, E-14071 Córdoba, Spain}
\email[]{ajcuesta@uco.es}

\author{Jailson Alcaniz}
\affiliation{Observatório Nacional, Rua General José Cristino, 77, São Cristóvão, 20921-400, Rio de Janeiro, RJ, Brazil}
\email[]{alcaniz@on.br}

\author{Gabriel Rodrigues}
\affiliation{Observatório Nacional, Rua General José Cristino, 77, São Cristóvão, 20921-400, Rio de Janeiro, RJ, Brazil}
\affiliation{Departamento de Física, Universidad de Córdoba, Campus Universitario de Rabanales, Ctra. N-IV Km. 396, E-14071 Córdoba, Spain}
\email[]{gabrielrodrigues@on.br}

\author{Miguel Aparicio Resco}
\affiliation{Universidad Europea de Madrid. Campus de Villaviciosa de Odón. C/ Tajo, s/n. Urb. El Bosque, Villaviciosa de Odón, 28670 Madrid, Spain}
\email[]{miguel.aparicio@universidadeuropea.es}

\author{Antonio López Maroto}
\affiliation{Departamento de Física Teórica and Instituto de Física de Partículas y del Cosmos (IPARCOSUCM), Universidad Complutense de Madrid, 28040 Madrid, Spain}
\email[]{maroto@ucm.es}

\author{Manuel Masip}
\affiliation{CAFPE and Departamento de Física Teórica y del Cosmos, Universidad de Granada, E18071 Granada, Spain}
\email[]{masip@ugr.es}

\author{Jamerson G. Rodrigues}
\affiliation{Observatório Nacional, Rua General José Cristino, 77, São Cristóvão, 20921-400, Rio de Janeiro, RJ, Brazil}
\email[]{jamersonrodrigues@on.br}

\author{Felipe B. M. dos Santos}
\affiliation{Observatório Nacional, Rua General José Cristino, 77, São Cristóvão, 20921-400, Rio de Janeiro, RJ, Brazil}
\email[]{fbmsantos@on.br}

\author{Javier de Cruz Pérez}
\affiliation{Departamento de Física, Universidad de Córdoba, Campus Universitario de Rabanales, Ctra. N-IV Km. 396, E-14071 Córdoba, Spain}
\email[]{jdecruz@uco.es}

\author{Jorge Enrique García-Farieta}
\affiliation{Departamento de Física, Universidad de Córdoba, Campus Universitario de Rabanales, Ctra. N-IV Km. 396, E-14071 Córdoba, Spain}
\email[]{jorge.farieta@uco.es}

\author{Raul Abramo}
\affiliation{Departamento de Física Matemática, Instituto de Física, Universidade de São Paulo, Rua do Matão, 1371, CEP 05508-090, São Paulo, Brazil}
\email[]{abramo@fma.if.usp.br}

\author{Narciso Benitez}
\affiliation{Independent Researcher}
\email[]{txitxo.benitez@gmail.com}

\author{Silvia Bonoli}
\affiliation{Donostia International Physics Center (DIPC), Manuel Lardizabal Ibilbidea, 4, San Sebastián, Spain}
\affiliation{Centro de Estudios de Física del Cosmos de Aragón (CEFCA), Plaza San Juan, 1, E-44001, Teruel, Spain}
\email[]{silvia.bonoli@dipc.org}

\author{Saulo Carneiro}
\affiliation{Observatório Nacional, Rua General José Cristino, 77, São Cristóvão, 20921-400, Rio de Janeiro, RJ, Brazil}
\email[]{saulocarneiro@on.br}

\author{Javier Cenarro}
\affiliation{Centro de Estudios de Física del Cosmos de Aragón (CEFCA), Plaza San Juan, 1, E-44001, Teruel, Spain}
\email[]{cenarro@cefca.es}

\author{David Cristóbal-Hornillos}
\affiliation{Centro de Estudios de Física del Cosmos de Aragón (CEFCA), Plaza San Juan, 1, E-44001, Teruel, Spain}
\email[]{dchornillos@gmail.com}

\author{Renato Dupke}
\affiliation{Observatório Nacional, Rua General José Cristino, 77, São Cristóvão, 20921-400, Rio de Janeiro, RJ, Brazil}
\email[]{rdupke@gmail.com}

\author{Alessandro Ederoclite}
\affiliation{Centro de Estudios de Física del Cosmos de Aragón (CEFCA), Plaza San Juan, 1, E-44001, Teruel, Spain}
\email[]{aederocl.astro@gmail.com}

\author{Antonio Hernán-Caballero}
\affiliation{Centro de Estudios de Física del Cosmos de Aragón (CEFCA), Plaza San Juan, 1, E-44001, Teruel, Spain}
\email[]{ahernan@cefca.es}

\author{Carlos Hernández–Monteagudo}
\affiliation{Instituto de Astrofísica de Canarias, C/ Vía Láctea, s/n, E-38205, La Laguna, Tenerife, Spain}
\affiliation{Universidad de La Laguna, Avda Francisco Sánchez, E-38206, San Cristóbal de La Laguna, Tenerife, Spain}
\email[]{chm@iac.es}

\author{Carlos López-Sanjuan}
\affiliation{Centro de Estudios de Física del Cosmos de Aragón (CEFCA), Plaza San Juan, 1, E-44001, Teruel, Spain}
\email[]{clsj@cefca.es}

\author{Antonio Marín-Franch}
\affiliation{Centro de Estudios de Física del Cosmos de Aragón (CEFCA), Plaza San Juan, 1, E-44001, Teruel, Spain}
\email[]{amarin@cefca.es}

\author{Claudia Mendes de Oliveira}
\affiliation{Departamento de Astronomia, Instituto de Astronomia, Geofísica e Ciências Atmosféricas, Universidade de São Paulo, São Paulo, Brazil}
\email[]{claudia.oliveira@iag.usp.br}

\author{Mariano Moles}
\affiliation{Centro de Estudios de Física del Cosmos de Aragón (CEFCA), Plaza San Juan, 1, E-44001, Teruel, Spain}
\email[]{moles@cefca.es}

\author{Laerte Sodré Jr.}
\affiliation{Departamento de Astronomia, Instituto de Astronomia, Geofísica e Ciências Atmosféricas, Universidade de São Paulo, São Paulo, Brazil}
\email[]{laerte.sodre@iag.usp.br}

\author{Keith Taylor}
\affiliation{Instruments4, 4121 Pembury Place, La Canada Flintridge, CA 91011, U.S.A.}
\email[]{kt.astro@gmail.com}

\author{Jesús Varela}
\affiliation{Centro de Estudios de Física del Cosmos de Aragón (CEFCA), Plaza San Juan, 1, E-44001, Teruel, Spain}
\email[]{jvarela@cefca.es}

\author{Héctor Vázquez Ramió}
\affiliation{Centro de Estudios de Física del Cosmos de Aragón (CEFCA), Plaza San Juan, 1, E-44001, Teruel, Spain}
\email[]{hvr@cefca.es}

\begin{abstract}
Fisher-matrix forecasts are presented for the cosmological surveys of the
Javalambre Physics of the Accelerating Universe Astrophysical Survey (J-PAS)
and the Subaru Prime Focus Spectrograph (PFS). The wide, low-redshift
coverage of J-PAS and the high-density, high-redshift mapping of PFS are
strongly complementary: combining the two reduces marginalized
uncertainties on all primary parameters compared with either survey
individually. Adding the joint J-PAS+PFS data to next-generation CMB
measurements from the Simons Observatory (SO) and \textsc{LiteBird} yields an expected precision
of $\sigma(\sum m_\nu)=0.017\,$eV in the
$\Lambda$CDM$+\sum m_\nu+N_{\rm eff}$ framework, sufficient to disfavour the
inverted neutrino hierarchy at $2.34\,\sigma$ if the true mass sum equals
the normal–ordering minimum.  Motivated by recent DESI results, we also
forecast within a $w_0w_a$CDM$+\sum m_\nu+N_{\rm eff}$ cosmology, adopting
the DESI\,DR2 best-fit values ($w_0=-0.758$, $w_a=-0.82$) as fiducial.
The combination CMB+J-PAS+PFS then delivers
$\sigma(w_0)=0.044$ and $\sigma(w_a)=0.18$, corresponding to a
$5.1\,\sigma$ preference for a time-varying dark‐energy equation of state.
These findings show that J-PAS and PFS, especially when coupled with
Stage-IV CMB observations, will provide competitive tests of neutrino
physics and the dynamics of cosmic acceleration.
\end{abstract}


\section{Introduction}

Oscillation experiments involving atmospheric and solar neutrinos have unequivocally demonstrated that neutrinos are massive fermions with three distinct mass eigenstates, $m_1$, $m_2$, and $m_3$ \citep{wolfensteinNeutrinoOscillationsMatter1978,collaborationEvidenceOscillationAtmospheric1998,ahmadDirectEvidenceNeutrino2002}.  
Although these experiments cannot determine the absolute mass scale, they provide precise measurements of the mass–squared splittings,
$\Delta m_{ij}^{2} \equiv m_i^{2}-m_j^{2}$, with the most up-to-date constraints  
$\Delta m_{21}^{2}=7.49^{+0.19}_{-0.19}\times10^{-5}\,{\rm eV}^{2}$ and  
$\Delta m_{31}^{2}=+2.513^{+0.021}_{-0.019}\times10^{-3}\,{\rm eV}^{2}$ or  
$\Delta m_{32}^{2}=-2.484^{+0.020}_{-0.020}\times10^{-3}\,{\rm eV}^{2}$  
\citep{estebanNuFit60UpdatedGlobal2024}.  
If the lightest eigenstate is assumed to be massless, these splittings translate into the following lower bounds on the sum of the neutrino masses,  
\begin{equation}
     \mv \equiv \sum_{i=1}^{3} m_i > 0.06\,\mathrm{eV}\qquad (m_1<m_2\ll m_3),
     \label{eq:NO}
\end{equation}
for the normal ordering (NO), and  
\begin{equation}
     \mv > 0.10\,\mathrm{eV}\qquad (m_3\ll m_1<m_2),
     \label{eq:IO}
\end{equation}
for the inverted ordering (IO).  
Determining $\mv$ therefore offers a direct route to discriminating between these two hierarchies.

Direct kinematic searches based on tritium $\beta$ decay, exemplified by the KATRIN experiment \citep{collaborationKATRINNextGeneration2001}, currently set the tightest $90\%$ confidence upper limit of $\mv<1.35\,\mathrm{eV}$ \citep{akerDirectNeutrinomassMeasurement2024}, which is still far from the sensitivity required to resolve the ordering.

Cosmology provides a complementary—and potentially far more sensitive—avenue for measuring $\mv$ \citep{huWeighingNeutrinosGalaxy1998,lesgourguesMassiveNeutrinosCosmology2006}.  
Massive neutrinos free-stream out of potential wells on scales below their characteristic free-streaming length, suppressing the growth of matter perturbations and imprinting a scale-dependent reduction in the small-scale matter power spectrum (see Sec.\,\ref{cosmo_effect}).  
Galaxy redshift surveys can detect this signature, but precise constraints are hampered by a well-known parameter degeneracy between $\mv$ and the dark-energy (DE) equation-of-state parameter \citep{hannestadNeutrinoMassesDark2005,lorenzDistinguishingNeutrinosTimevarying2017,liuUpdateConstraintsNeutrino2020}.

The most recent cosmological bounds for neutrino masses employ the baryon acoustic oscillations (BAO) measured from the second data release (DR2) of the Dark Energy Spectroscopic Instrument (DESI) \citep{collaborationDESIDR2Results2025,elbersConstraintsNeutrinoPhysics2025}.  
Combined with Planck CMB and Atacama Cosmology Telescope (ACT) CMB-lensing data, DESI DR2 yields $\mv<0.064\,\mathrm{eV}$ (95\% C.L.) within the $\Lambda$CDM framework, a value already approaching the NO lower bound in Eq.\,(\ref{eq:NO}).  
In a $\wcdm$ context, in which the equation of state of dark energy is parametrized as $w(a) =w_0+w_a(1-a)$ with two parameters $w_0$ and $w_a$, the limit relaxes to $\mv<0.16\,\mathrm{eV}$ (95\% C.L.), consistent with both Eqs.\,(\ref{eq:NO}) and (\ref{eq:IO}).  
These limits rest on the minimal physical prior $\mv>0$, which forces the posterior to peak at the boundary.  
To quantify the impact of this prior, \citet{elbersNegativeNeutrinoMasses2025} introduced an effective neutrino-mass parameter, $\mveff$, identical to $\mv$ for positive values but allowed to assume negative values under a broad uniform prior.  
Several recent analyses find that cosmological data exhibit a mild preference for $\mveff<0$ \citep{craigNo$n$sGood2024,greenCosmologicalPreferenceNegative2024,naredo-tueroLivingEdgeCritical2024,elbersNegativeNeutrinoMasses2025,elbersConstraintsNeutrinoPhysics2025}.  
In particular, DESI DR2 favors $\mveff=-0.101\,\mathrm{eV}$ in $\Lambda$CDM, corresponding to a $3.0\sigma$ tension with the positive NO bound \citep{collaborationDESIDR2Results2025,elbersConstraintsNeutrinoPhysics2025}.

Stage-IV large scale structure surveys promise to improve these constraints even further.  
Besides DESI, imminent programs include the Javalambre Physics of the Accelerating Universe Astrophysical Survey (J-PAS; \citealt{benitezJPASJavalambrePhysicsAccelerated2014}) and the Prime Focus Spectrograph (PFS; \citealt{takadaExtragalacticScienceCosmology2014}).  
J-PAS will map low-redshift tracers with unparalleled spectral resolution, whereas PFS will observe emission-line galaxies (ELGs) out to $z\sim2.4$, supplying high-density sampling at intermediate and high redshift (see Sec.\,\ref{surveys}).  
Their complementary redshift coverage motivates a joint analysis: the low-$z$ leverage from J-PAS breaks degeneracies that afflict the high-$z$ PFS data, enabling simultaneously tighter constraints on $\mv$ and DE parameters.

In this work we present Fisher-matrix forecasts for the combined J-PAS and PFS surveys and compare them with analogous forecasts for the full five-year DESI data set.  
Our methodology is tailored to demonstrate how the synergy between J-PAS and PFS can mitigate parameter degeneracies across a broad redshift range.

The remainder of this paper is organized as follows.  
Section~\ref{data} describes the impact of massive neutrinos on large-scale structure, summarizes the survey specifications of J-PAS and PFS, and details our modeling of the observed galaxy power spectrum.  
Section~\ref{fisher_formalism} presents the Fisher-matrix formalism together with its numerical implementation.  
The resulting parameter forecasts are reported in Section~\ref{result}, and our main conclusions are summarized in Section~\ref{conclusion}.

\section{Model and Data} \label{data}

\subsection{Effects of Neutrinos on Large-Scale Structure of the Universe} \label{cosmo_effect}

We begin by briefly reviewing how massive neutrinos alter both the homogeneous expansion history and the
growth of large-scale structure (LSS). In a spatially flat $\Lambda$CDM+$\nu$ cosmology,
the Hubble expansion rate is
\begin{equation}
\begin{aligned}
  H(z)=H_0 \{&\,\Omega_{\mathrm r,0}(1+z)^4
                 +\Omega_{\mathrm{cb},0}(1+z)^3
                 +\Omega_{\Lambda,0} \\
                 &+\Omega_{\nu,0}(1+z)^{3}\exp\left[\int_0^z\frac{3w_\nu(z')}{1+z'}\dd z'\right]\}^{1/2},
\end{aligned}
\label{eq:Hz}
\end{equation}
where $\Omega_{i,0}\equiv\rho_{i,0}/\rho_{\mathrm{cr}}$ is the present-day energy-density fraction of
component $i=\{\mathrm r,\mathrm{cb},\Lambda,\nu\}$, with
$\mathrm{cb}$ denoting cold plus baryonic matter.
The neutrino equation of state,
$w_\nu(z)\equiv P_\nu/\rho_\nu$, evolves from the relativistic value $1/3$ at early times
to the non-relativistic value $0$ after the transition redshift
\begin{equation}
  z_{\mathrm{nr}}\simeq1890\!\left(\frac{m_\nu}{\mathrm{eV}}\right),
\label{eq:znr}
\end{equation}
so that $\rho_\nu\propto(1+z)^4$ for $z\gg z_{\mathrm{nr}}$ and
$\rho_\nu\propto(1+z)^3$ for $z\ll z_{\mathrm{nr}}$.
Because massive neutrinos free-stream out of dark-matter potential wells below their
characteristic length scale, they suppress the growth of structure on small scales,
imprinting a scale-dependent reduction in the matter power spectrum amplitude that is
approximately proportional to the neutrino mass fraction
$f_\nu\equiv\Omega_{\nu,0}/\Omega_{m,0}$.

\begin{figure}[htb!]
\centering
\includegraphics[width=0.47\textwidth]{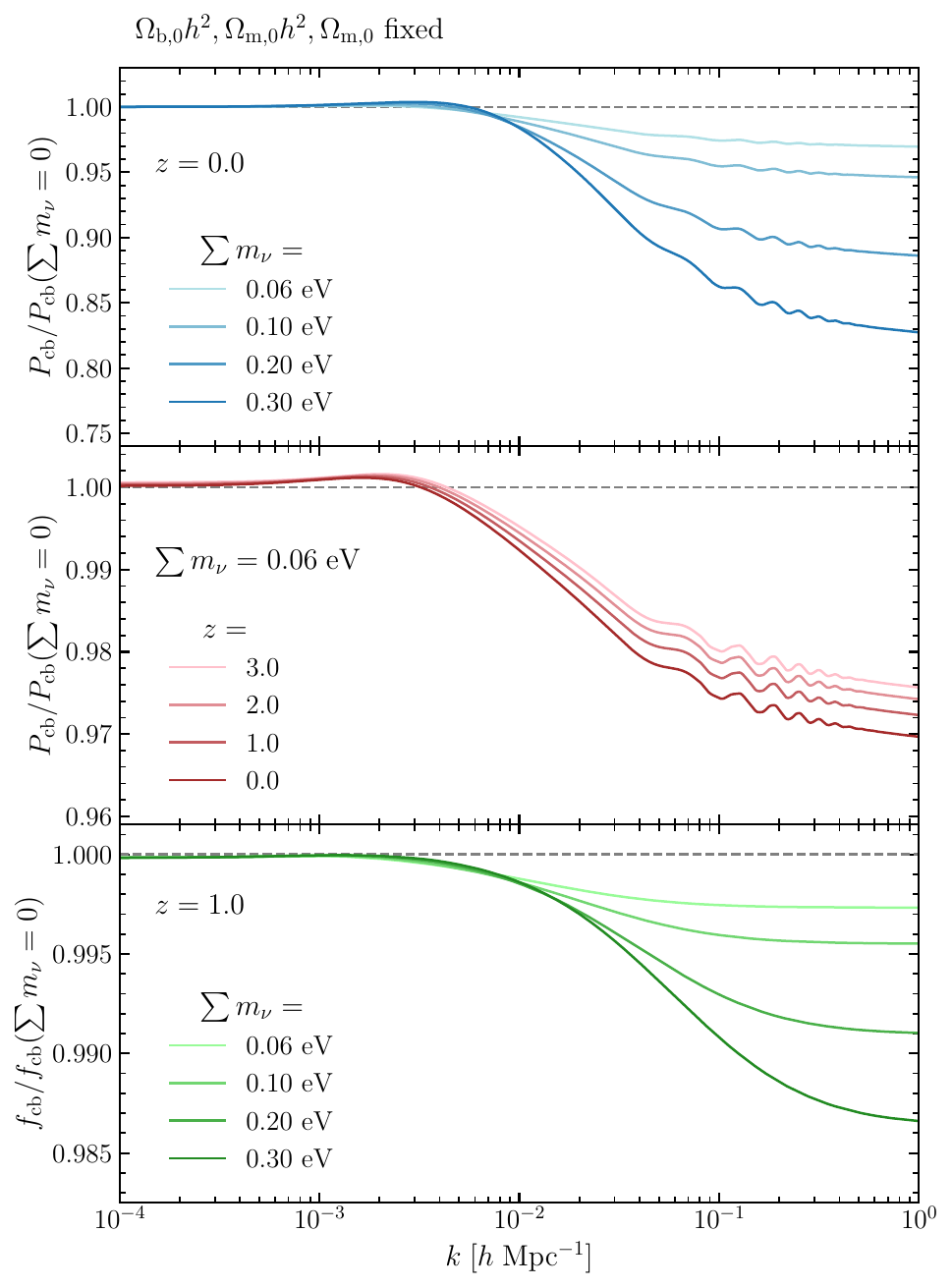}
\caption{Top: Linear cold-plus-baryonic matter power spectrum at $z=0$ for varying $\sum m_\nu$, normalized to the massless case. Middle: Power spectrum for $\sum m_\nu = 0.06$\,eV relative to $\sum m_\nu = 0$ at different redshifts. Bottom: Linear growth rate of CDM+baryons at $z=1$ normalized to the massless case.}
\label{suppression}
\end{figure}

While neutrinos are still relativistic their energy density simply tracks that of the photon bath,
\begin{equation}
  \Omega_{\nu,0}^{\mathrm{rel}}
    = \frac{7}{8}\left(\frac{4}{11}\right)^{4/3}N_{\mathrm{eff}}\,
      \Omega_{\mathrm r,0},
  \label{eq:onu_rel}
\end{equation}
where $N_{\mathrm{eff}}$ is the effective number of relativistic species and the numerical prefactor
encodes the standard ratio of neutrino to CMB temperatures after $e^\pm$ annihilation. Note that we adopt the instant decoupling approximation here.
Once the neutrinos become non-relativistic their energy density redshifts like matter and is fixed by the
sum of their masses \citep{frousteyUniverseMeVEra2022},
\begin{equation}
  \Omega_{\nu,0}^{\mathrm{nr}}
    = \frac{\sum m_\nu}{93.12\,h^{2}\,\mathrm{eV}}.
  \label{eq:onu_nr}
\end{equation}

Large thermal velocities allow massive neutrinos to free-stream out of dark-matter potential wells,
suppressing the growth of structure on scales smaller than the comoving free-streaming length
\citep{lesgourguesMassiveNeutrinosCosmology2006}.  A convenient measure of this scale is the
free-streaming wavenumber,
\begin{equation}
  k_{\mathrm{FS}}(z)\simeq
    0.82\,h\,\mathrm{Mpc}^{-1}\,
    \frac{H(z)}{H_0(1+z)^{2}}
    \left(\frac{m_i}{1\,\mathrm{eV}}\right),
  \label{eq:kfs}
\end{equation}
where $m_i$ denotes the mass of a single neutrino eigenstate.  Modes with
$k\gtrsim k_{\mathrm{FS}}$ experience a scale-dependent
suppression whose amplitude is roughly proportional to the neutrino mass fraction
$f_\nu\equiv\Omega_{\nu,0}/\Omega_{m,0}$.

\begin{figure}
\centering
\includegraphics[width=0.47\textwidth]{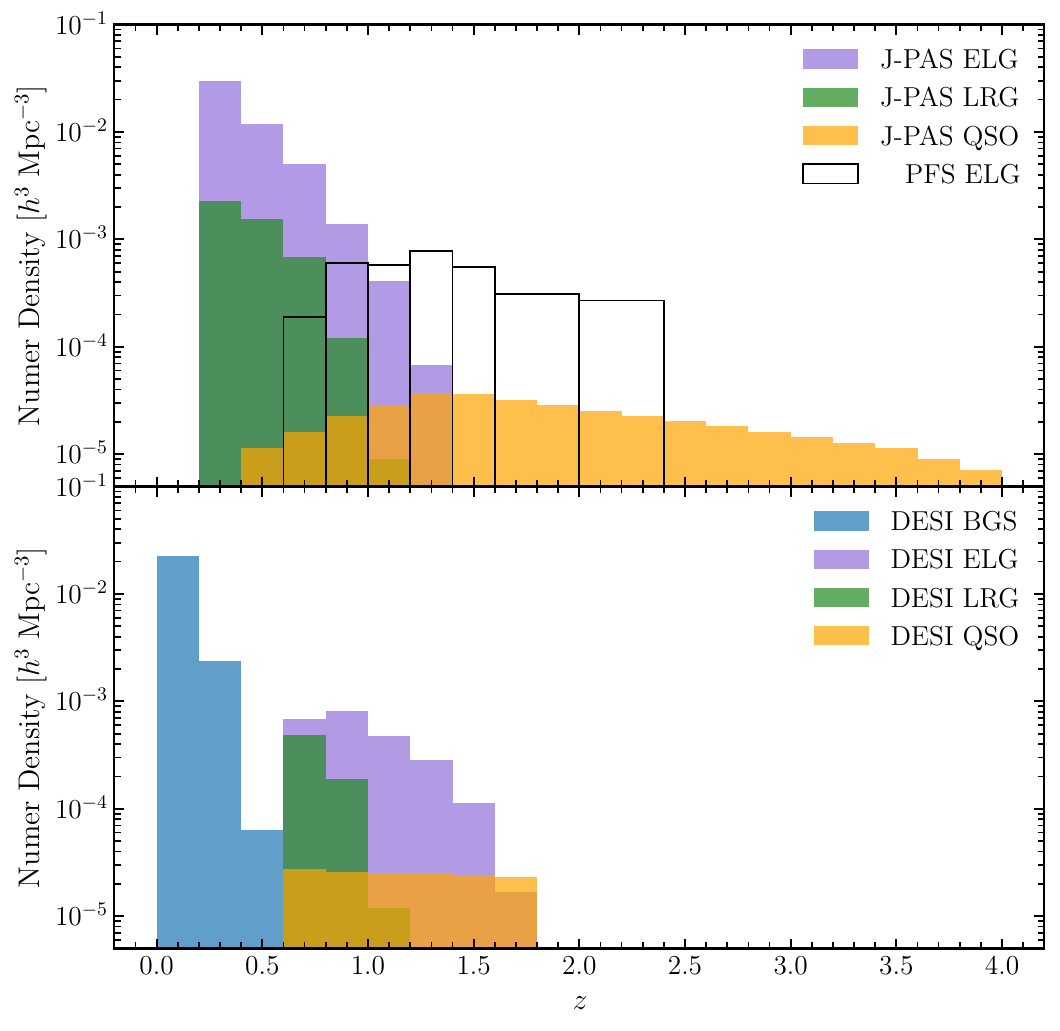}
\caption{Volume number density of targets in redshift bins. Top: J-PAS (shaded) and PFS (solid line). Bottom: DESI BGS, ELG, LRG, and QSO (shaded).}
\label{density}
\end{figure}

Figure~\ref{suppression} illustrates the influence of massive neutrinos on the cold plus baryon density field as a function of scale and redshift. Here, we keep the values of ($\Omega_\mathrm{b,0}h^2,\Omega_\mathrm{m,0}h^2,\Omega_\mathrm{m,0}$) fixed, while the CDM energy density ($\Omega_\mathrm{c,0}h^2$) decreases as the neutrino energy density ($\Omega_\mathrm{\nu,0}h^2$) increases, in order to preserve the total matter density. The top panel displays the ratio of linear cold-plus-baryonic matter power spectrum $P_{\mathrm{cb}}(k;\sum m_\nu)/P_{\mathrm{cb}}(k;0)$ at $z=0$ for four representative values of the total neutrino mass. A larger $\sum m_\nu$ produces a deeper suppression at wavenumbers above the free streaming scale, and the depth is approximately proportional to the neutrino mass fraction $f_\nu\equiv\Omega_{\nu,0}/\Omega_{m,0}$. The middle panel shows the same ratio for a fixed $\sum m_\nu = 0.06\,\mathrm{eV}$ at several redshifts; the suppression weakens at earlier times because the neutrinos are less nonrelativistic and the free streaming scale is larger. The bottom panel presents the corresponding fractional reduction in the linear logarithmic growth rate, $f_{\mathrm{cb}}(k)\equiv \mathrm d\ln \delta_{\mathrm{cb}}/\mathrm d\ln a$, at $z=1$. This panel demonstrates that the dynamical effect closely matches the static suppression observed in the power spectrum.

\subsection{The J-PAS and PFS Surveys} \label{surveys}

The Javalambre Physics of the Accelerating Universe Astrophysical Survey (J-PAS; \citealt{benitezJPASJavalambrePhysicsAccelerated2014,bonoliMiniJPASSurveyPreview2021}) is a wide field, narrow band imaging program carried out with the 2.55\,m Javalambre Survey Telescope at the Observatorio Astrofísico de Javalambre in Spain. Using a dedicated set of 54 contiguous narrow band filters, together with three broad bands, J-PAS will map about $8500\ \mathrm{deg}^2$ of the northern sky and deliver photometric redshifts with a precision of $\sigma_z \simeq 0.003(1+z)$, comparable to low resolution spectroscopy. The survey is expected to detect roughly $9\times10^{7}$ galaxies and quasars, including luminous red galaxies, emission line galaxies, and quasistellar objects, which allows for a multi-tracer  analysis for improving the constraint on cosmological parameters \citep{seljakExtractingPrimordialNonGaussianity2009,mcdonaldHowEvadeSample2009}. The multi-tracer technique has already been applied in the analysis of observational data from the SDSS-IV eBOSS cosmology survey \citep{wangClusteringSDSSIVExtended2020,zhaoCompletedSDSSIVExtended2021,zhaoMultitracerAnalysisEBOSS2024}.
The corresponding comoving number densities and tomographic redshift bins of J-PAS survey adopted in this work are listed in Table~\ref{table:j-pas} \citep{apariciorescoJPASForecastsDark2020,Costa_2019,Salzano_2021,Figueruelo_2021}.

\setlength{\tabcolsep}{12pt} 
\setlength{\tabcolsep}{10pt}
\begin{deluxetable*}{cccc}
\tablecaption{Number density of J-PAS targets in each redshift bin. The survey covers $8500\ \mathrm{deg}^2$ with redshift precision $\sigma_{z}=0.003(1+z)$. All number densities are in units of $10^{-4}\,h^{3}\,\mathrm{Mpc}^{-3}$.\label{table:j-pas}}
\tablewidth{\textwidth}
\tablehead{
\colhead{Redshift bin} & \colhead{LRG} & \colhead{ELG} & \colhead{QSO}
}
\startdata
$0.2<z<0.4$  & 22.66 & 295.86 & 0.045 \\
$0.4<z<0.6$  & 15.63 & 118.11 & 0.114 \\
$0.6<z<0.8$  & 6.88  & 50.21  & 0.161 \\
$0.8<z<1.0$  & 1.20  & 13.80  & 0.227 \\
$1.0<z<1.2$  & 0.09  & 4.12   & 0.286 \\
$1.2<z<1.4$  & 0     & 0.67   & 0.360 \\
$1.4<z<1.6$  & 0     & 0      & 0.360 \\
$1.6<z<1.8$  & 0     & 0      & 0.321 \\
$1.8<z<2.0$  & 0     & 0      & 0.286 \\
$2.0<z<2.2$  & 0     & 0      & 0.255 \\
$2.2<z<2.4$  & 0     & 0      & 0.227 \\
$2.4<z<2.6$  & 0     & 0      & 0.203 \\
$2.6<z<2.8$  & 0     & 0      & 0.181 \\
$2.8<z<3.0$  & 0     & 0      & 0.161 \\
$3.0<z<3.2$  & 0     & 0      & 0.143 \\
$3.2<z<3.4$  & 0     & 0      & 0.128 \\
$3.4<z<3.6$  & 0     & 0      & 0.114 \\
$3.6<z<3.8$  & 0     & 0      & 0.091 \\
$3.8<z<4.0$  & 0     & 0      & 0.072 \\
\enddata
\end{deluxetable*}

The Prime Focus Spectrograph (PFS; \citealt{takadaExtragalacticScienceCosmology2014}) is mounted at the prime focus of the 8.2\,m Subaru Telescope on Maunakea, Hawaii, and will carry out a stage IV spectroscopic redshift survey of emission line galaxies. PFS is designed to reach a redshift accuracy of $\sigma_z \simeq 0.0007(1+z)$ for targets in the range $0.8 < z < 2.4$. The baseline program will obtain spectra for roughly four million galaxies distributed across $1464\ \mathrm{deg}^2$. The comoving number densities and tomographic redshift bins for PFS are listed in Table~\ref{table:pfs}.

\setlength{\tabcolsep}{20pt} 
\begin{deluxetable}{ccc}
\tablecaption{PFS survey specifications including redshift bins, fiducial ELG bias, and number densities. The sky coverage is $1464\ \mathrm{deg}^2$, redshift precision is $\sigma_{z}=0.0007(1+z)$. Number densities are in units of $10^{-4}\,h^{3}\,\mathrm{Mpc}^{-3}$.\label{table:pfs}}
\tablehead{
\colhead{Redshift bin} & \colhead{$b_{\rm fid}$} & \colhead{ELG}
}
\startdata
$0.6<z<0.8$  & 1.18 & 1.9 \\
$0.8<z<1.0$  & 1.26 & 6.0 \\
$1.0<z<1.2$  & 1.34 & 5.8 \\
$1.2<z<1.4$  & 1.42 & 7.8 \\
$1.4<z<1.6$  & 1.50 & 5.5 \\
$1.6<z<2.0$  & 1.62 & 3.1 \\
$2.0<z<2.4$  & 1.78 & 2.7 \\
\enddata
\end{deluxetable}

For completeness we also include the Dark Energy Spectroscopic Instrument
(DESI; \citealt{collaborationDESIExperimentPart2016}).  DESI started its five year
main survey in 2021 and will obtain spectra for more than
$3\times10^{7}$ galaxies and quasars across
$14\,000\ \mathrm{deg}^2$ of extragalactic sky, qualifying it as a Stage IV
spectroscopic program.  The target classes and their redshift dependent
comoving number densities adopted in this work are listed in
Table~\ref{table:desi}.  Figure~\ref{density} compares the comoving
number densities of the tracers used by J-PAS, PFS, and DESI.

For multi-tracer surveys such as J-PAS and DESI we describe the linear bias of each tracer population with  
\begin{equation}\label{bias}
     b(z)=
     \begin{cases}
          b_0/D_\mathrm{cb}(z), & \text{BGS, LRG, ELG},\\[4pt]
          0.53 + 0.289\,(1+z)^2, & \text{QSO},
     \end{cases}
\end{equation}
where \(b_0 = 1.34\) for the bright-galaxy sample (BGS), \(b_0 = 1.7\) for luminous red galaxies (LRGs), and \(b_0 = 0.84\) for emission-line galaxies (ELGs).  
Here \(D_\mathrm{cb}(z)\) is the linear growth factor of the combined cold-dark-matter and baryon fluid evaluated on large scales, where neutrino free–streaming can be neglected.  
We ignore any residual \(k\)-dependence in \(D_\mathrm{cb}(z)\) and in the bias itself; for the cold-plus-baryon power spectrum this assumption is accurate at the percent level over the wavenumber range used in our analysis \citep{euclidcollaborationEuclidPreparationSensitivity2024}.

Because the footprints of J-PAS and PFS are almost non-overlapping, their galaxy samples are statistically independent.  When we combine the two surveys we therefore add their Fisher matrices directly.

\setlength{\tabcolsep}{10pt} 
\begin{deluxetable*}{ccccc}
\tablecaption{DESI survey number densities in redshift bins for each tracer. The sky coverage is $14000\ \mathrm{deg}^2$. Redshift precision is $\sigma_{z}=0.0005(1+z)$ for BGS, LRG, and ELG, and $\sigma_{z}=0.001(1+z)$ for QSOs. Number densities are in units of $10^{-4}\,h^{3}\,\mathrm{Mpc}^{-3}$.\label{table:desi}}
\tablehead{
\colhead{Redshift bin} & \colhead{BGS} & \colhead{LRG} & \colhead{ELG} & \colhead{QSO}
}
\startdata
$0.0<z<0.2$  & 224.00 & 0     & 0     & 0     \\
$0.2<z<0.4$  & 24.00  & 0     & 0     & 0     \\
$0.4<z<0.6$  & 0.63   & 0     & 0     & 0     \\
$0.6<z<0.8$  & 0      & 4.87  & 6.91  & 0.275 \\
$0.8<z<1.0$  & 0      & 1.91  & 8.19  & 0.260 \\
$1.0<z<1.2$  & 0      & 0.118 & 4.77  & 0.255 \\
$1.2<z<1.4$  & 0      & 0     & 2.82  & 0.250 \\
$1.4<z<1.6$  & 0      & 0     & 1.12  & 0.240 \\
$1.6<z<1.8$  & 0      & 0     & 0.168 & 0.230 \\
\enddata
\end{deluxetable*}

\subsection{Modeling of Galaxy Survey Observables}

Following \citet{euclidcollaborationEuclidPreparationSensitivity2024} and \citet{casasEuclidValidationMontePython2023}, we describe the anisotropic galaxy power spectrum in redshift space by
\begin{equation} \label{galaxy_power_spectrum}
\begin{aligned}
P_\mathrm{g}(k,\mu,z) &= \left[ \frac{b\sigma_{8,\mathrm{cb}}(z) + f_{\mathrm{cb}}(k,z)\sigma_{8,\mathrm{cb}}(z)\mu^2}{1 + f_{\mathrm{cb,fid}}^2(k,z)k^2\mu^2\sigma^2_{\mathrm{p,fid}}} \right]^2 \\
&\quad \times \frac{P_\mathrm{dw}(k,\mu,z)}{\sigma_{8,\mathrm{cb}}^2(z)}\mathcal{F}_z(k,\mu,z),
\end{aligned}
\end{equation}
where $\mu \equiv \hat{\mathbf k}\!\cdot\!\hat{\mathbf n}$ is the cosine of the angle between the wavevector and the line of sight.

\begin{itemize}
  \item $f_{\mathrm{cb}}(k,z) \equiv \dd\ln \delta_{\mathrm{cb}} / \dd\ln a$ is the scale-dependent logarithmic growth rate of the combined cold-dark-matter and baryon (cb) density field.
  \item $\sigma_{8,\mathrm{cb}}(z)$ denotes the root-mean-square fluctuation amplitude of that field within spheres of radius $8\,h^{-1}\mathrm{Mpc}$.
  \item $P_\mathrm{dw}(k,\mu,z)$ is a ``de-wiggled’’ template that damps baryon-acoustic-oscillation (BAO) features,
        \begin{equation}
        \begin{aligned}
        P_\mathrm{dw}(k,\mu,z) =& P_\mathrm{cb}(k,z)\,e^{-g_\mu k^2} \\
        &+ P_\mathrm{nw}(k,z)\left[ 1 - e^{-g_\mu k^2} \right],
        \end{aligned}
        \end{equation}
        with $P_\mathrm{nw}(k)$ obtained by applying a Savitzky–Golay filter to smooth the BAO wiggles \citep{boyleDeconstructingNeutrinoMass2018}.
  \item The damping scale $g_\mu = \sigma_{\mathrm{v,fid}}^2\!\left[1 - \mu^2 + \mu^2 (1 + f_{\mathrm{cb,fid}})^2 \right]$ depends on the fiducial one-dimensional velocity dispersion $\sigma_{\mathrm{v,fid}}\simeq\sigma_{\mathrm{p,fid}}$, where
        \begin{equation}
        \sigma^2_{\mathrm{p}}(z) = \frac{1}{6\pi^2} \int \dd k\,P_{\mathrm{cb}}(k,z).
        \end{equation}
  \item Redshift-measurement uncertainties are captured by the factor
        $\mathcal{F}_z(k,\mu,z) = \exp\!\bigl[-k^2\mu^2\sigma_r^2(z)\bigr]$ with $\sigma_r(z)=c\,\sigma_{0,z}(1+z)/H(z)$.
\end{itemize}

To incorporate Alcock–Paczynski (AP) distortions \citep{alcockEvolutionFreeTest1979}, the power spectrum measured under the fiducial cosmology is rescaled as
\begin{equation} \label{AP}
P_\mathrm{obs}(k_\mathrm{fid},\mu_\mathrm{fid},z) = \frac{1}{q_\parallel q_\perp^2}\,
P_\mathrm{g}\!\left( \frac{k_\mathrm{fid}G}{q_\perp}, \frac{\mu_\mathrm{fid}\,q_\perp}{G\,q_\parallel}, z \right) + P_\mathrm{shot}(z),
\end{equation}
where
\begin{align}
q_\parallel(z) &= \frac{H(z)}{H_{\mathrm{fid}}(z)}, \qquad
q_\perp(z)   = \frac{D_{\mathrm{A,fid}}(z)}{D_{\mathrm{A}}(z)}, \\[4pt]
G &= \sqrt{1 + \mu^2_{\mathrm{fid}}\!\left(\frac{q_\perp^2}{q_\parallel^2} - 1\right)}.
\end{align}

Shot noise is modeled by
\begin{equation} \label{shot_noise}
P_{\mathrm{shot}}(z) = \frac{1}{\bar{n}(z)} + p_{\mathrm{s},z}(z),
\end{equation}
where $\bar{n}(z)$ is the comoving number density of tracers and the residual term $p_{\mathrm{s},z}$ is set to zero in our fiducial analysis.

The linear cold-plus-baryon spectrum $P_\mathrm{cb}(k,z)$ is computed with \texttt{CAMB} \citep{lewisEfficientComputationCMB2000a} and evolved into the nonlinear regime using \texttt{HMcode2020} \citep{meadHMcode2020ImprovedModelling2021}, which provides percent-level accuracy in cosmologies that include massive neutrinos and outperforms the older \texttt{Halofit} model \citep{takahashiRevisingHalofitModel2012}.

\section{Fisher Forecasting Methodology}\label{fisher_formalism}

This section introduces the Fisher information framework we use to quantify the constraining power of forthcoming galaxy redshift surveys and cosmic microwave background data. Section~\ref{subsec:FMzsurvey} explains how we build the Fisher matrix for the galaxy power spectrum, while Section~\ref{subsec:FMcmb} outlines the parallel treatment of primary CMB temperature and polarization observables. The adopted parameter set, fiducial values, priors, and numerical details are gathered in Section~\ref{subsec:parasetting}.

\subsection{Fisher Matrix for Galaxy Surveys} \label{subsec:FMzsurvey}

The Fisher matrix for galaxy redshift surveys \citep{tegmarkKarhunenLoeveEigenvalueProblems1997} is given by
\begin{equation}\label{fisher2}
     F_{ij}(z)=\int_{-1}^{1}\dd\mu\int_{k_{\mathrm{min}}}^{k_{\mathrm{max}}}\frac{\dd k\,k^2V_\mathrm{fid}}{4\pi^2}\mathrm{Tr}\left(\frac{\partial\mathbb{C}}{\partial\theta_i}\mathbb{C}^{-1}\frac{\partial\mathbb{C}}{\partial\theta_j}\mathbb{C}^{-1}\right)\,,
\end{equation}
where $\theta_i$ are the model parameters, and $V_\mathrm{fid}(z)$ is the comoving volume of the redshift bin in the fiducial cosmology. The covariance matrix for multiple tracers is
\begin{equation}
     \mathbb{C}=
     \begin{bmatrix}
          C^{\rm XX} & C^{\rm XY} & \cdots \\
          C^{\rm YX} & C^{\rm YY} & \cdots \\
          \vdots & \vdots & \ddots \\
     \end{bmatrix}\,,
\end{equation}
where ${\rm X, Y}, \dots$ denote different tracer types. The minimum wavenumber is set to $k_{\mathrm{min}}=0.007\,h\,\mathrm{Mpc}^{-1}$ for J-PAS \citep{apariciorescoJPASForecastsDark2020}, and $k_{\mathrm{min}}=1\times10^{-4}\,h\,\mathrm{Mpc}^{-1}$ for both PFS \citep{takadaExtragalacticScienceCosmology2014} and DESI. The maximum wavenumber is fixed at $k_{\mathrm{max}}=0.25\,h\,\mathrm{Mpc}^{-1}$ for all three surveys. Our choice of $k_{\mathrm{max}}=0.25\,h\,\mathrm{Mpc}^{-1}$ is comparable to what is often referred to as the ``pessimistic" scale cut in \cite{casasEuclidValidationMontePython2023} and \cite{euclidcollaborationEuclidPreparationSensitivity2024}.

\subsubsection{The single-tracer case}

If only one tracer is observed (e.g., for PFS), then
\begin{equation}
\mathbb{C}=C^{\rm XX}=P_{\mathrm{obs}}\,,
\end{equation}
where $P_{\mathrm{obs}}$ is the observed galaxy power spectrum with noise as given in Eq.\,(\ref{AP}). Equation~(\ref{fisher2}) then reduces to \citep{abramoWhyMultitracerSurveys2013}
\begin{equation}
     \begin{aligned}
          F_{ij}(z) = &\frac{1}{8\pi^2}\int_{-1}^1\dd\mu\int_{k_{\mathrm{min}}}^{k_{\mathrm{max}}}\dd k\,k^2\frac{\partial\ln\mathcal{P}}{\partial\theta_i}\left(\frac{\mathcal{P}}{\mathcal{P}+1}\right)^2  \\ &\times\frac{\partial\ln\mathcal{P}}{\partial\theta_j}V_\mathrm{fid}\,,
     \end{aligned}
\end{equation}
where $\mathcal{P}=\bar{n}(z)P_{\mathrm{obs}}$ is the effective power spectrum.

\subsubsection{The multi-tracer case}

If multiple tracers are used (e.g., in J-PAS or DESI), Eq.\,(\ref{fisher2}) can be rewritten as \citep{abramoWhyMultitracerSurveys2013}
\begin{equation}
     \begin{aligned}
          F_{ij}(z) = & \sum_{\rm XY}\frac{1}{16\pi^2}\int_{-1}^1\dd\mu\int_{k_{\mathrm{min}}}^{k_{\mathrm{max}}}\dd k\,k^2\frac{\partial\ln\mathcal{P}_{\rm X}}{\partial\theta_i}                                                                              \\
                     & \times\left[\delta_{\rm XY}\frac{\mathcal{P}_{\rm X}\mathcal{P}}{1+\mathcal{P}}+\frac{\mathcal{P}_{\rm X}\mathcal{P}_{\rm Y}(1-\mathcal{P})}{(1+\mathcal{P})^2}\right]\frac{\partial\ln\mathcal{P}_{\rm Y}}{\partial\theta_j}V_\mathrm{fid}\,,
     \end{aligned}
\end{equation}
where ${\rm X}$ and ${\rm Y}$ index the tracer types, $\mathcal{P} = \sum_{\rm X} \mathcal{P}_{\rm X}$, and $\delta_{\rm XY}$ is the Kronecker delta.

The total Fisher matrix is then obtained by summing over redshift bins:
\begin{equation}
     F_{ij}=\sum_z F_{ij}(z)
\end{equation}

\subsection{Fisher Matrix for CMB Observables}  \label{subsec:FMcmb}

\begin{deluxetable*}{lcCccCC}
\tablecaption{\fx{Survey specifications for the LiteBIRD and SO experiments.}}\label{table:cmb}
\tablecolumns{7}
\tablewidth{0pt}
\tablehead{
\colhead{Component} & 
\colhead{Frequency} & 
\colhead{FWHM} & 
\colhead{Multipole Range} & 
\colhead{$f_{\mathrm{sky}}$} & 
\colhead{$\Delta T$} & 
\colhead{$\Delta P$} \\
\colhead{} & 
\colhead{[GHz]} & 
\colhead{[arcmin]} & 
\colhead{} & 
\colhead{} & 
\colhead{[$\mu$K\,arcmin]} & 
\colhead{[$\mu$K\,arcmin]}
}
\startdata
LiteBIRD (low-$\ell$)  & 140 & 31   & $2 \leq \ell \leq 40$     & 0.7  &  4.1  & 5.8  \\
\multirow{2}{*}{\fx{SO (high-$\ell$)}}       & \fx{93}  & \fx{2.2}  & \fx{$41 \leq \ell \leq 3000$}  & \fx{0.61} &  \fx{5.3}  & \fx{7.5}  \\
                       & \fx{145} & \fx{1.4}  &~\fx{$41 \leq \ell \leq 3000$ } & \fx{0.61} &  \fx{6.6}  & \fx{9.3}  \\
LiteBIRD (high-$\ell$) & 140 & 31   & $41 \leq \ell \leq 1350$  & 0.09 &  4.1  & 5.8  \\
\enddata
\tablecomments{The table summarizes the angular resolution (via FWHM), multipole coverage, sky fraction, and instrumental noise for both temperature ($\Delta T$) and polarization ($\Delta P$) channels of the LiteBIRD and \fx{SO} experiments.}
\end{deluxetable*}

\begin{deluxetable*}{cccccccccc}
\tablecaption{Fiducial values of cosmological parameters used in our forecast.\label{table:fiducial}}
\tablehead{
\colhead{$h$} & 
\colhead{$\Omega_\mathrm{m,0}$} & 
\colhead{$\Omega_\mathrm{b,0}$} & 
\colhead{$\sigma_8$} & 
\colhead{$n_\mathrm{s}$} & 
\colhead{$\sum m_\nu$ [eV]} & 
\colhead{$N_\mathrm{eff}$} & 
\colhead{$w_0$} & 
\colhead{$w_a$} & 
\colhead{$\tau$}
}
\startdata
0.67 & 0.32 & 0.049 & 0.81 & 0.96 & 0.06 & 3.044 & $-1$ & 0 & 0.054 \\
\enddata
\end{deluxetable*}

We also consider combining the forecast results from next-generation CMB experiments, 
\fx{such as the Simons Observatory (SO; \citealt{collaborationSimonsObservatoryScience2019, collaborationSimonsObservatoryScience2025})}
and LiteBIRD \citep{matsumuraMissionDesignLiteBIRD2014, collaborationProbingCosmicInflation2023}. The CMB observables used in our analysis include the lensed temperature (TT), polarization (EE), and their cross power spectra (TE), represented as 
\begin{equation}
\mathbf{C}_\ell = \left[C_\ell^\mathrm{TT}, C_\ell^\mathrm{EE}, C_\ell^\mathrm{TE}\right]^\mathrm{T}.
\end{equation}
The Fisher matrix for the CMB observables is given by \citep{abazajianCMBS4ScienceBook2016}
\begin{equation}
F_{ij}=\sum_\ell\frac{\partial\mathbf{C}_\ell^{\mathbf{T}}}{\partial\theta_i}\mathbb{C}_\ell^{-1}\frac{\partial\mathbf{C}_\ell}{\partial\theta_j}\,,
\end{equation}
where $\mathbb{C}_\ell$ is the covariance matrix of the CMB observables. It is computed as
\begin{equation}
\begin{aligned}
\mathbb{C}_\ell(C_\ell^{\alpha\beta},C_\ell^{\gamma\delta}) &= \frac{1}{(2\ell+1)f_\mathrm{sky}}\left[(C_\ell^{\alpha\gamma}+N_\ell^{\alpha\gamma})(C_\ell^{\beta\delta}+N_\ell^{\beta\delta})\right. \\
& \left.+(C_\ell^{\alpha\delta}+N_\ell^{\alpha\delta})(C_\ell^{\beta\gamma}+N_\ell^{\beta\gamma})\right],
\end{aligned}
\end{equation}
where $f_\mathrm{sky}$ denotes the fraction of the sky covered by the CMB survey, and $N_\ell$ represents the instrumental noise. Any combination of two from $\alpha$, $\beta$, $\gamma$, and $\delta$ corresponds to one of the observable combinations: TT, EE, or TE. The noise power spectrum $N_\ell$ (for either temperature or polarization) is calculated as \citep{pogosianTrackingDarkEnergy2005}
\begin{equation}
N_\ell^{-1}=\sum_{c}\left\{\sigma_c^2\exp\left[\frac{\ell(\ell+1)\Theta^2_{\mathrm{FWHM},c}}{8\ln2}\right]\right\}^{-1}\,,
\end{equation}
where $\sigma_c$ is the instrumental noise level in $\mu\mathrm{K\,arcmin}$, and $\Theta_{\mathrm{FWHM},c}$ is the full width at half maximum (FWHM) of the beam in arcminutes. We adopt the survey specifications 
\fx{for the SO Large Aperture Telescope (LAT) mid-frequency (MF) band from \citet{collaborationSimonsObservatoryScience2025} and LiteBIRD from \citet{brinckmannPromisingFutureRobust2019}, as summarized in Table~\ref{table:cmb}. We consider a combination of these two surveys consisting of LiteBIRD data for $\ell\leq40$, SO data for $\ell\geq41$ and additional LiteBIRD data for $\ell\geq41$ covering the sky region not covered by SO. This is similar to the combination of LiteBIRD and CMB-S4 considered in \citet{brinckmannPromisingFutureRobust2019}.}
 For temperature measurements, $\sigma_c = \Delta T$ and $N_\ell = N_\ell^\mathrm{TT}$; for polarization, $\sigma_c = \Delta P$ and $N_\ell = N_\ell^\mathrm{EE}$. We assume no cross-correlation in the noise, i.e., $N_\ell^\mathrm{TE} = 0$.

\subsection{Parameters and Numerical Calculations}  \label{subsec:parasetting}

The basic cosmological parameters in the $\Lambda$CDM model are given by
\begin{equation}
\{\theta_i^{\rm base}\} = \left\{h,\Omega_\mathrm{m,0},\Omega_\mathrm{b,0},\sigma_8,n_\mathrm{s}\right\}\,.
\end{equation}
We consider four extended cosmological scenarios, each with additional parameters:
\begin{itemize}
    \item $\Lambda$CDM + $\mv$: \\
    \hspace*{1.5em} $\{\theta_i\} = \{\theta_i^{\rm base}, \mv\}$
    \item $\Lambda$CDM + $\mv$ + $N_{\rm eff}$: \\
    \hspace*{1.5em} $\{\theta_i\} = \{\theta_i^{\rm base}, \mv, N_{\rm eff}\}$
    \item $w_0 w_a$CDM + $\mv$: \\
    \hspace*{1.5em} $\{\theta_i\} = \{\theta_i^{\rm base}, w_0, w_a, \mv\}$
    \item $w_0 w_a$CDM + $\mv$ + $N_{\rm eff}$: \\
    \hspace*{1.5em} $\{\theta_i\} = \{\theta_i^{\rm base}, w_0, w_a, \mv, N_{\rm eff}\}$
\end{itemize}

The fiducial values of the cosmological parameters used in our forecast are summarized in Table~\ref{table:fiducial}. They are consistent with the TT,TE,EE+lowE+lensing result of Planck 2018 \citep{collaborationPlanck2018Results2020}.

In the analysis of galaxy surveys, we include two additional nuisance parameters for each redshift bin $z$: $\left\{\ln(b\sigma_8)_z, p_{\mathrm{s},z}\right\}$, which appear in Eqs.~(\ref{galaxy_power_spectrum}) and (\ref{shot_noise}). These parameters are treated as independent and are appended to the parameter set $\{\theta_i\}$ for each cosmological scenario. Therefore, for $N_z$ redshift bins, the total number of free parameters becomes $(7 + 2N_z)$ for the $\Lambda$CDM + $\mv$ + $N_{\rm eff}$ model and $(9 + 2N_z)$ for the $w_0 w_a$CDM + $\mv$ + $N_{\rm eff}$ model.

\begin{figure*}[htb!]
    \centering
    \includegraphics[width=0.93\textwidth]{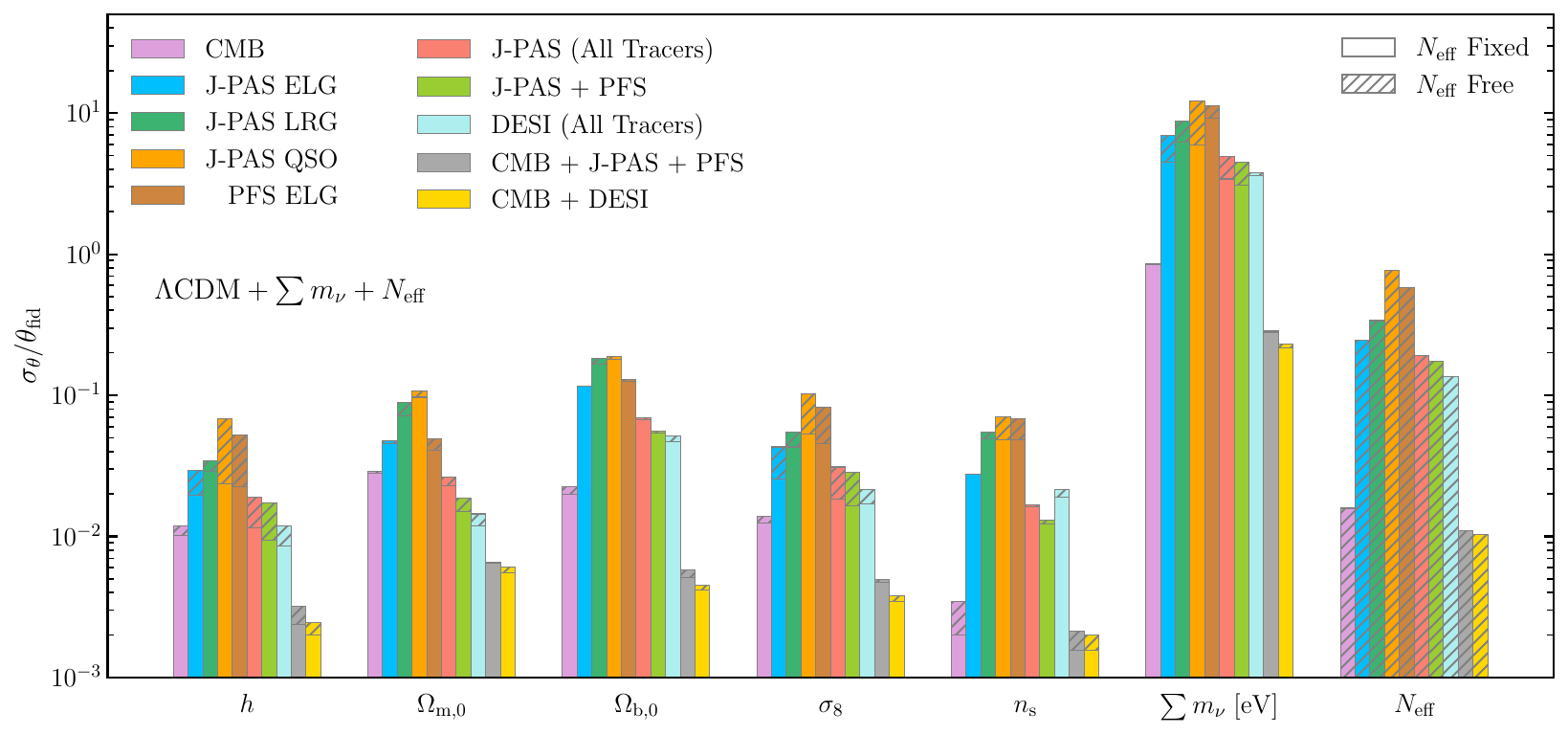}
    \includegraphics[width=0.93\textwidth]{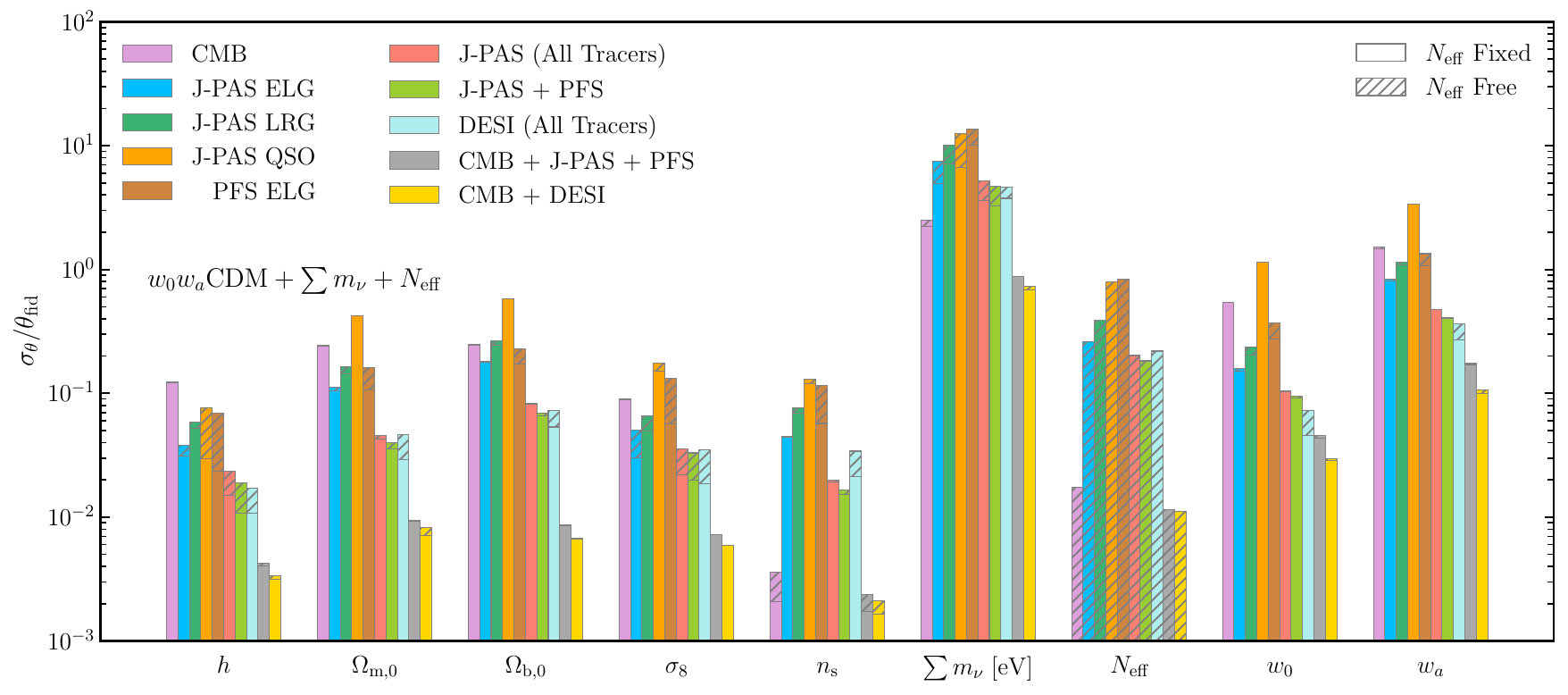}
    \caption{\fx{Upper: Fractional uncertainties on cosmological parameters for different data combinations in the $\Lambda$CDM$+\sum m_\nu$ (shaded) and $\Lambda$CDM$+\sum m_\nu+N_\mathrm{eff}$ (hatched) models; Lower: same as the upper figure but for the $w_0w_a$CDM$+\sum m_\nu+N_\mathrm{eff}$ model. For the parameters $w_0$ and $w_a$, the uncertainties $\sigma_{\theta}$ are presented.}}
    \label{barplot}
\end{figure*}

In the analysis of CMB observables, the optical depth to reionization $\tau$ is included as a free parameter and added to the parameter set $\{\theta_i\}$. Its fiducial value is listed in Table~\ref{table:fiducial}. Since $\tau$ is not a parameter of direct interest, it is always marginalized over in the final cosmological constraints. For galaxy surveys, $\tau$ is fixed at its fiducial value.

We compute partial derivatives in the Fisher formalism using central finite differences:
\begin{equation}
    \frac{\partial \mathcal{O}(\theta_i)}{\partial\theta_i} \approx \frac{\mathcal{O}(\theta_i + \Delta\theta_i) - \mathcal{O}(\theta_i - \Delta\theta_i)}{2\Delta\theta_i}\,,
\end{equation}
where the observable is $\mathcal{O}(\theta_i) = \mathcal{P}(\theta_i)$ for galaxy surveys and $\mathcal{O}(\theta_i) = \mathbf{C}_\ell(\theta_i)$ for CMB data. 
For the neutrino mass $\mv$, we adopt a step size of 10\% of its fiducial value. For $w_0$ and $w_a$, the step size is set to 0.01. For all other parameters, we use a step size of 1\% of the fiducial value. We adopt the same step size for the numerical derivatives as used in \cite{euclidcollaborationEuclidPreparationSensitivity2024}.

The derivatives with respect to the nuisance parameters in the galaxy survey analysis are computed analytically:
\begin{equation}
    \frac{\partial \ln \mathcal{P}(z)}{\partial p_{\mathrm{s},z}} = \frac{1}{P_{\mathrm{obs}}(z)},
\end{equation}
and
\begin{equation}
    \frac{\partial \ln \mathcal{P}(k,\mu,z)}{\partial \ln(b\sigma_8)_z} = \frac{2b(z)}{b(z) + f(k,z)\mu^2}.
\end{equation}

We developed a dedicated Fisher forecast code\footnote{\url{https://github.com/Striker-png/NeutrinoForecast_J-PAS_PFS}} to implement the analysis described in this section.

\section{Results} \label{result}

\begin{figure*}[htb!]
    \centering
    \includegraphics[width=0.93\textwidth]{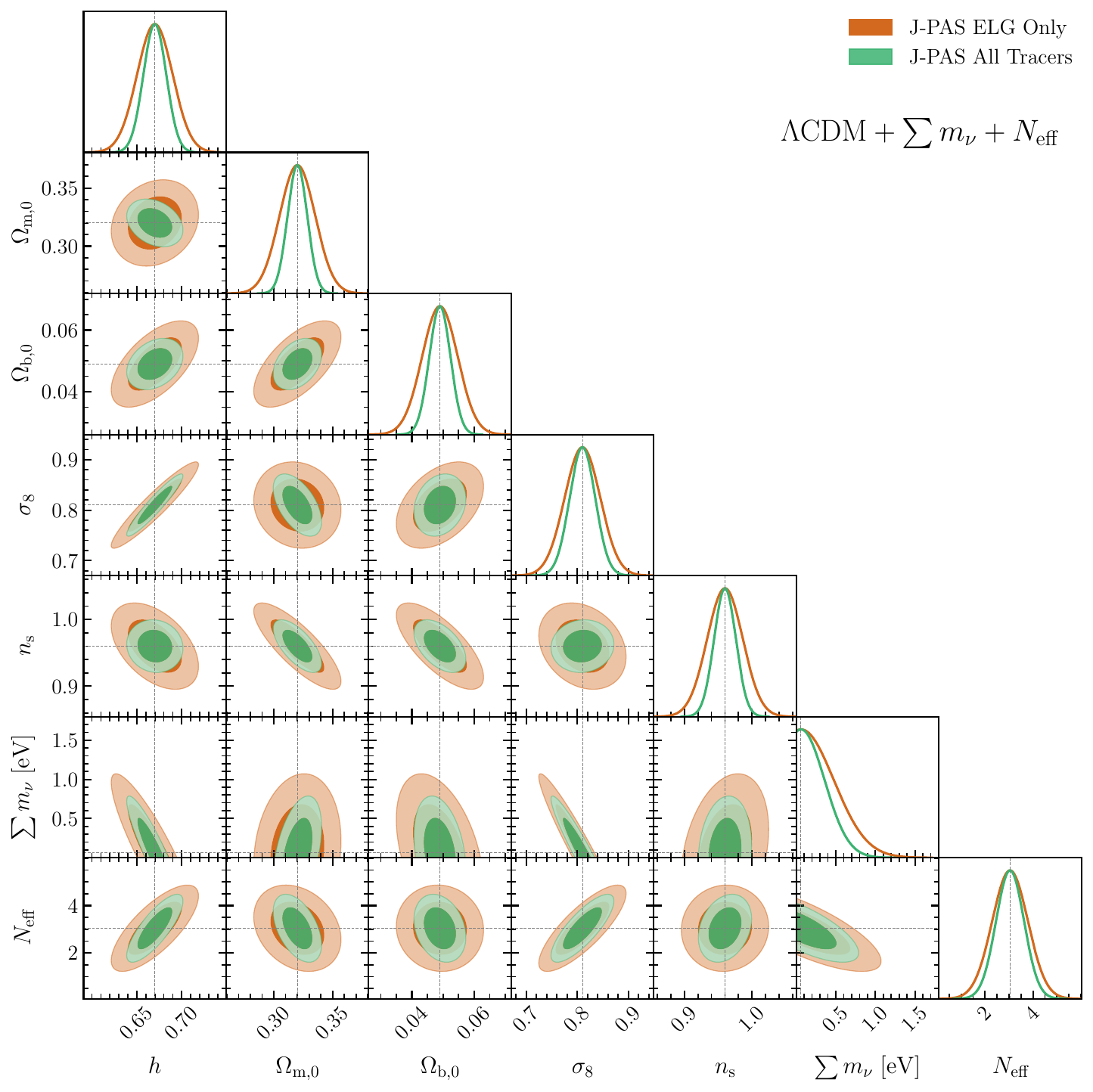}
    \caption{Marginalized 68\% and 95\% confidence contours for the $\Lambda$CDM$+\sum m_\nu+N_\mathrm{eff}$ model using J-PAS ELG only (red) and the combination of three tracers (green).}
    \label{lcdm_j-pas}
\end{figure*}

\begin{figure*}[htb!]
    \centering
    \includegraphics[width=0.93\textwidth]{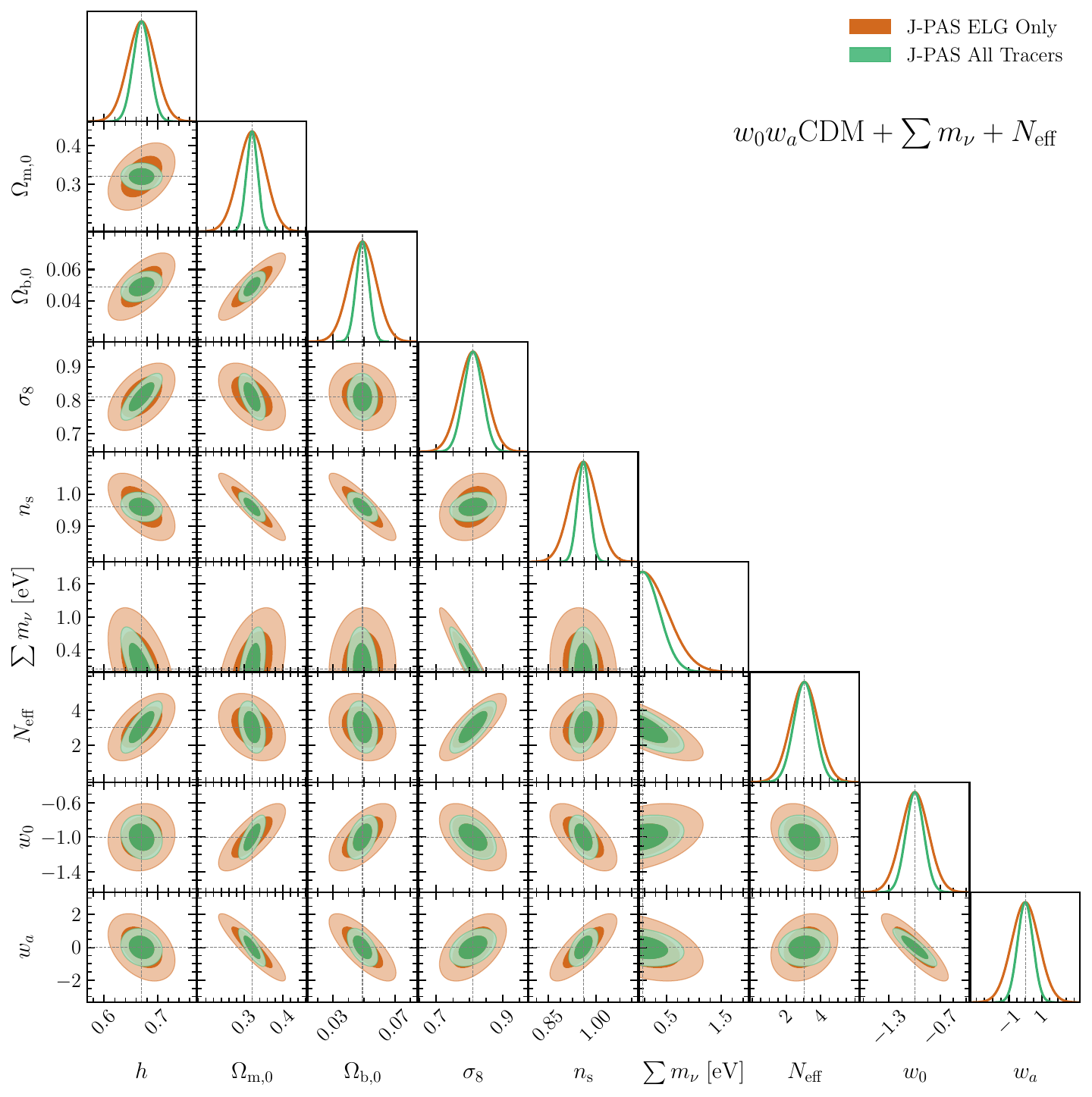}
    \caption{Same as Figure~\ref{lcdm_j-pas}, but for the $w_0w_a$CDM$+\sum m_\nu+N_\mathrm{eff}$ model.}
    \label{w0wacdm_j-pas}
\end{figure*}

\begin{figure*}[htb!]
    \centering
    \includegraphics[width=0.475\textwidth]{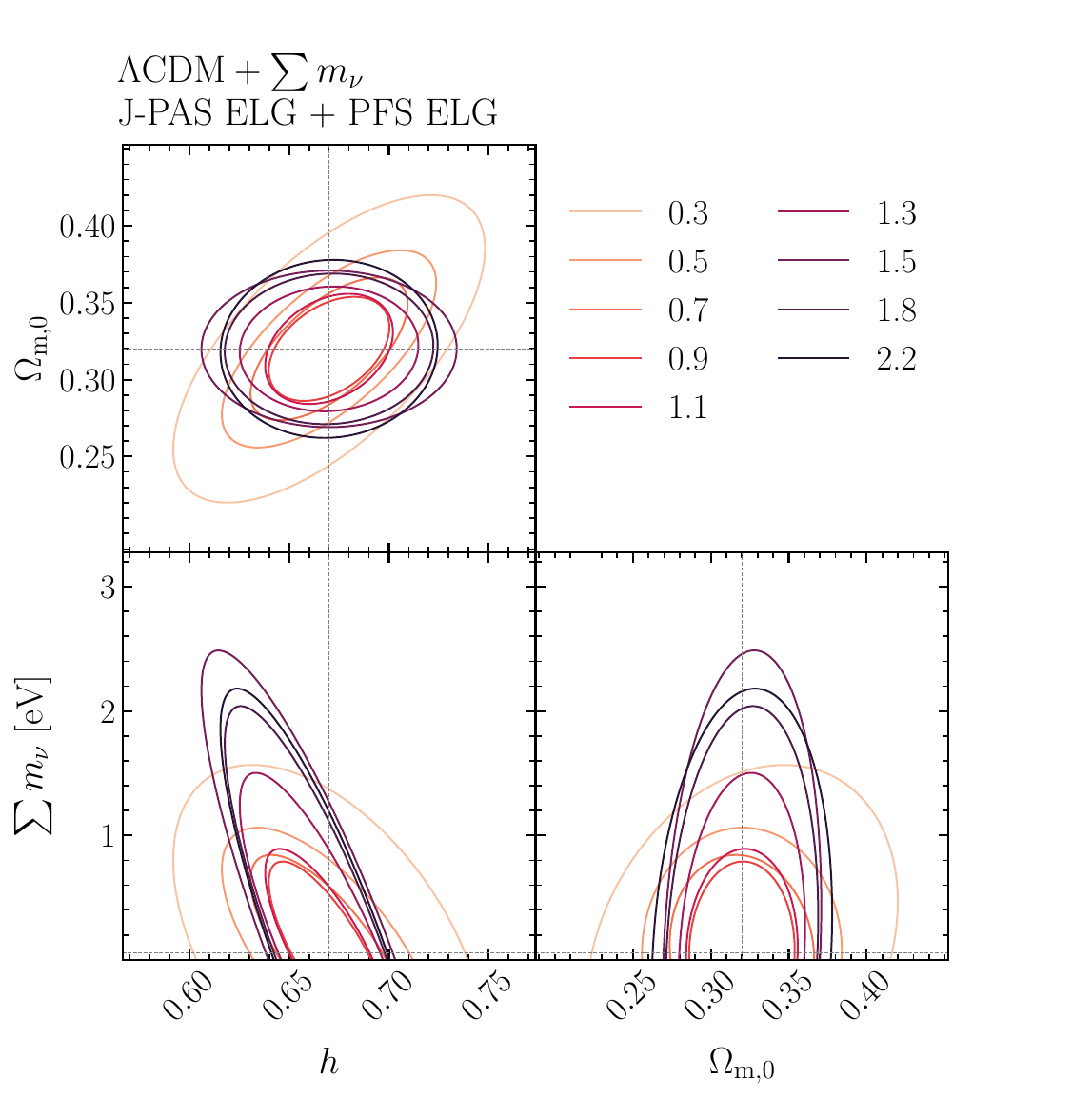}
    \includegraphics[width=0.47\textwidth]{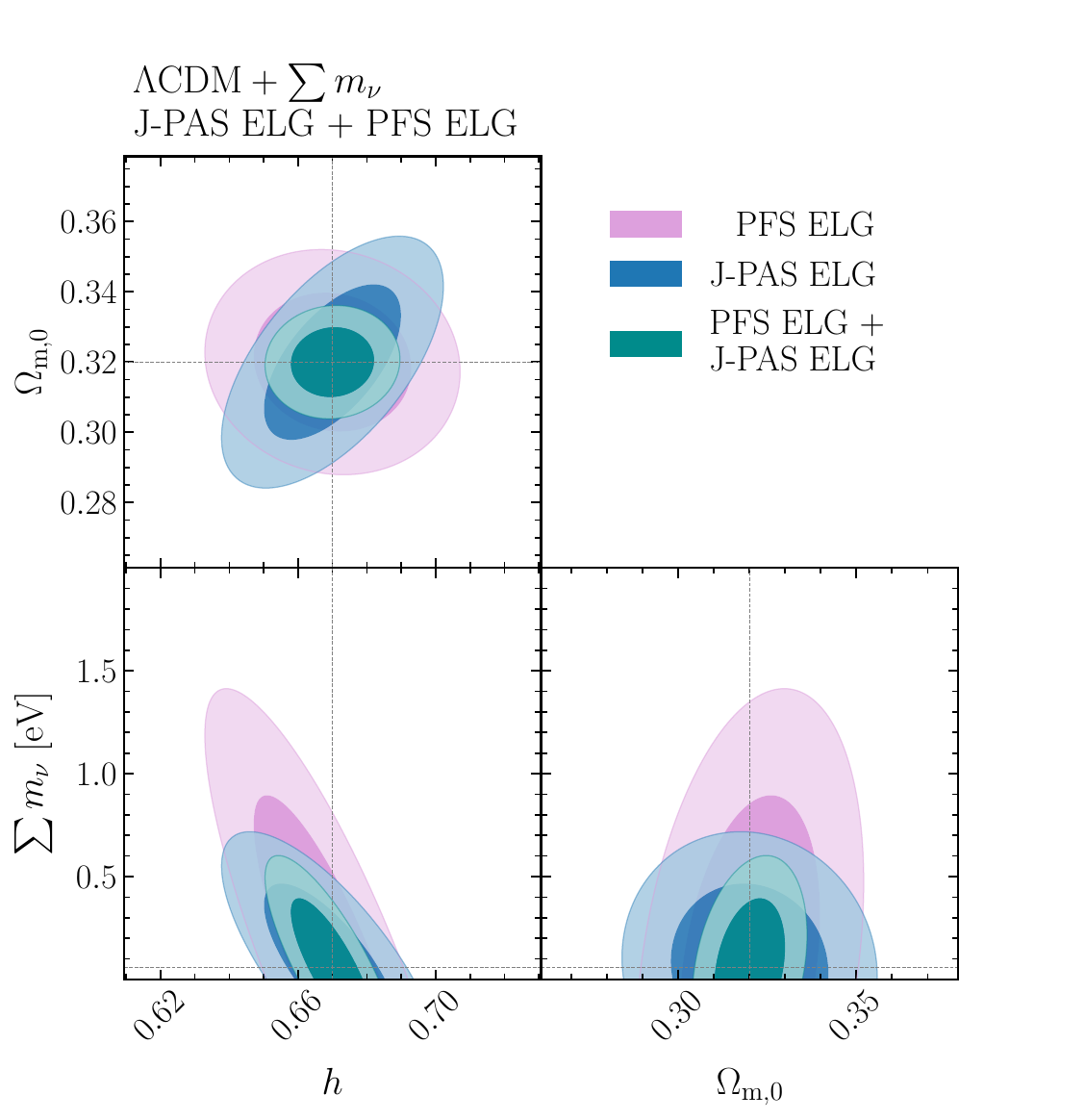}
    \caption{Left: 68\% confidence contours for ($h$, $\Omega_\mathrm{m,0}$, $\mv$)} in each redshift bin from J-PAS ELG + PFS ELG in the $\Lambda$CDM$+\sum m_\nu$ model. Right: Joint constraints from PFS ELG (magenta), J-PAS ELG (blue), and their combination (green).
    \label{lcdm_elg_z}
\end{figure*}

\begin{figure*}[htb!]
    \centering
    \includegraphics[width=0.475\textwidth]{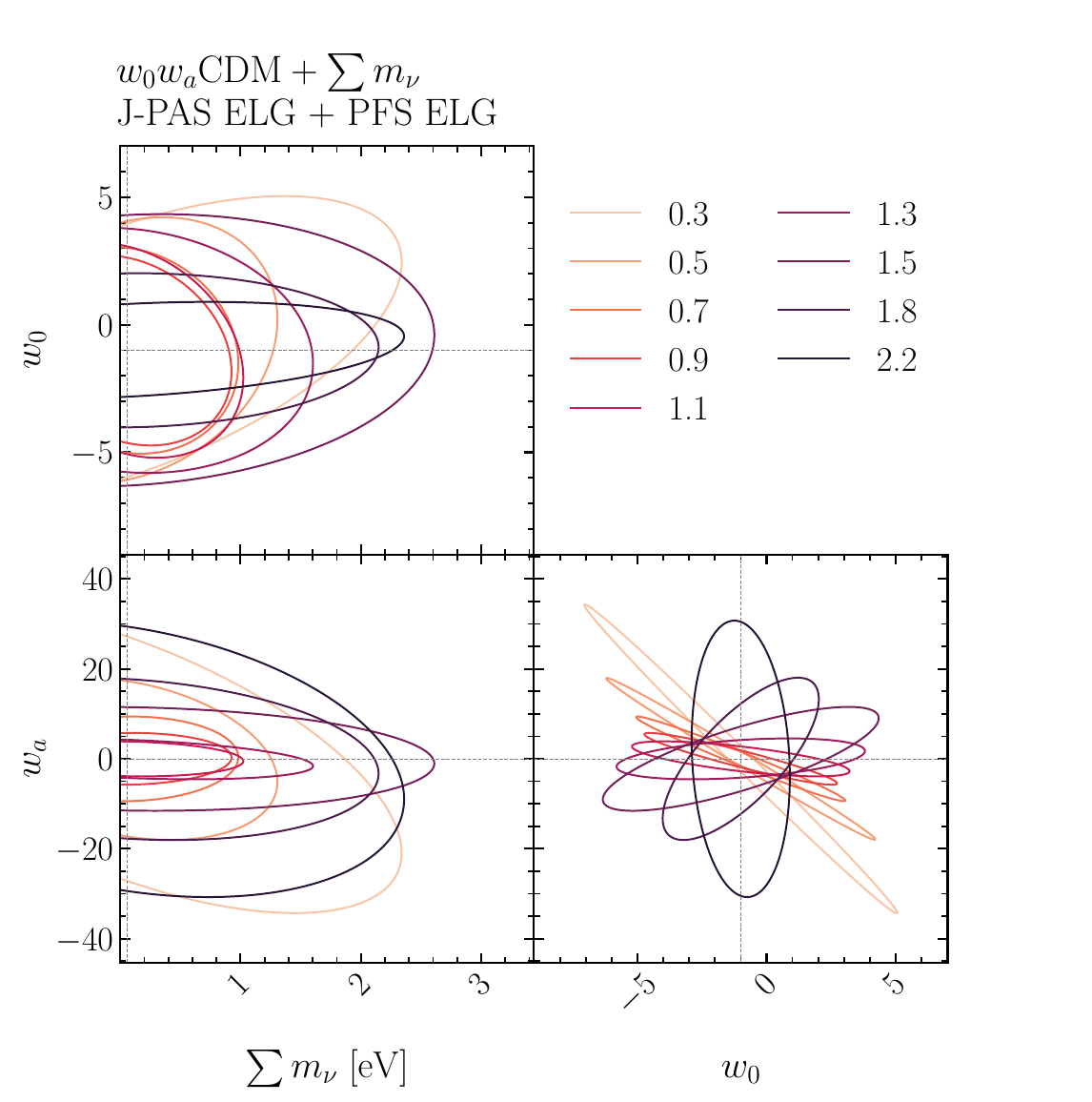}
    \includegraphics[width=0.47\textwidth]{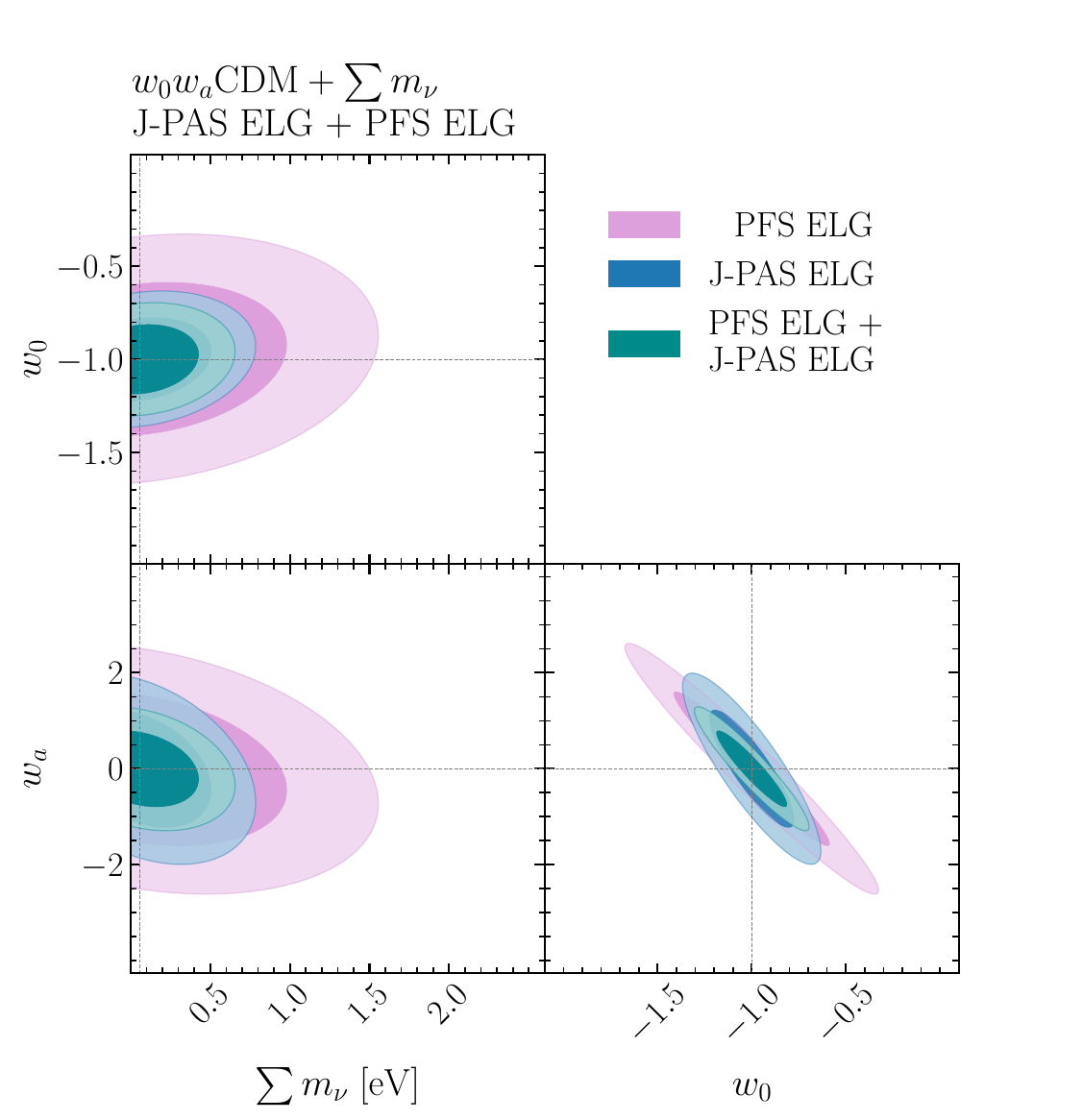}
    \caption{Same as Figure~\ref{lcdm_elg_z}, but for ($\sum m_\nu$, $w_0$, $w_a$) in the $w_0w_a$CDM$+\sum m_\nu$ model.}
    \label{w0wacdm_elg_z}
\end{figure*}

In this section we present forecasts for the marginalized uncertainties on cosmological parameters obtained from several combinations of data sets, summarized in Figures \ref{barplot}-\ref{de} and in Table \ref{tab:results}. Section~\ref{subsec:JPAS} summarizes the single tracer and multi tracer results derived from the J-PAS survey.  Section~\ref{subsec:z} investigates how the J-PAS emission-line galaxy sample complements the higher-redshift emission-line galaxies observed by PFS.  Section~\ref{subsec:joint} combines the galaxy survey information with temperature and polarization measurements of the cosmic microwave background to provide our tightest joint constraints.

\subsection{Single-Tracer and Multi-Tracer Results from J-PAS} \label{subsec:JPAS}

The upper part of Figure~\ref{barplot} displays the fractional uncertainties on cosmological parameters for the $\Lambda$CDM$+\sum m_\nu$ (shaded bars) and $\Lambda$CDM$+\sum m_\nu+N_\mathrm{eff}$ (hatched bars) scenarios. Results are shown for J-PAS ELG, LRG, and QSO samples individually, the combination of all three (multi-tracer), and their combination with PFS ELG. For comparison, we include the forecasts from DESI and CMB-only, as well as the combined constraints from CMB with J-PAS and PFS or with DESI (see Sec.\,\ref{subsec:joint} for the discussion of the combined results). The fixed-$N_\mathrm{eff}$ case yields tighter constraints than the case with free $N_\mathrm{eff}$, as expected.

Among individual J-PAS tracers, ELGs deliver the strongest constraints due to their high number density. Combining all three tracers significantly improves the results. Specifically, the multi-tracer analysis improves the constraints relative to ELG-only forecasts by factors of (1.6, 1.8, 1.7, 1.4, 1.6, 1.4, 1.3) for ($h$, $\Omega_\mathrm{m,0}$, $\Omega_\mathrm{b,0}$, $\sigma_8$, $n_\mathrm{s}$, $\sum m_\nu$, $N_\mathrm{eff}$), respectively. Different scenarios for redshift errors and number densities in the J-PAS survey are explored and discussed in \cite{rodriguesForecasts2025}.

Figure~\ref{lcdm_j-pas} shows marginalized 68\% and 95\% confidence contours for these parameters, further highlighting the improvement from combining tracers. Throughout this paper, in the contour plots we consistently present only the positive values of the neutrino mass.

We extend the analysis to the $w_0w_a$CDM$+\sum m_\nu+N_\mathrm{eff}$ model, in which the equation of state of dark energy is described as $w(a)=w_0+w_a(1-a)$ \citep{chevallierACCELERATINGUNIVERSESSCALING2001,linderExploringExpansionHistory2003},
in the lower part of Figure~\ref{barplot}, showing fractional uncertainties with $N_\mathrm{eff}$ fixed and free. ELGs again provide the best single-tracer performance, and the multi-tracer strategy yields further gains. The improvement factors over ELG-only are $(1.6, 2.4, 2.2, 1.4, 2.3, 1.5, 1.3, 1.5, 1.8)$ for ($h$, $\Omega_\mathrm{m,0}$, $\Omega_\mathrm{b,0}$, $\sigma_8$, $n_\mathrm{s}$, $\sum m_\nu$, $N_\mathrm{eff}$, $w_0$, $w_a$), respectively.

Figure~\ref{w0wacdm_j-pas} displays the marginalized contours for the nine parameters in this extended model, again confirming that combining tracers substantially improves parameter constraints.

\subsection{Redshift Complementarity from J-PAS and PFS} \label{subsec:z}

\begin{figure*}[htb!]
  \centering
  \includegraphics[width=0.93\textwidth]{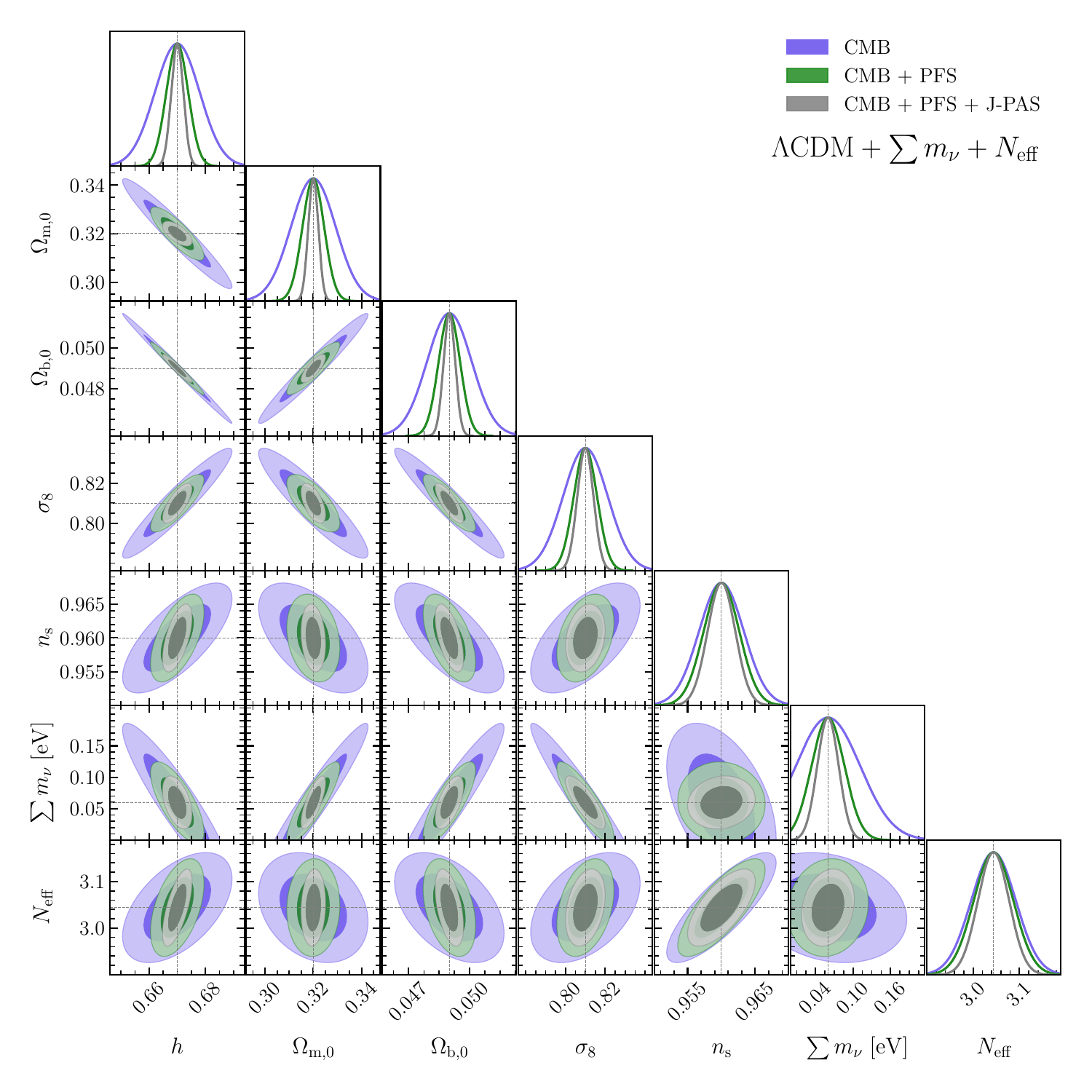}
  \caption{\fx{Marginalized 68\% (inner) and 95\% (outer) confidence contours for
    the $\Lambda$CDM$+\sum m_\nu+N_{\rm eff}$ cosmology obtained from CMB data
    alone (purple), CMB\,+\,PFS\,ELG (green), and the full combination
    CMB\,+\,PFS\,ELG\,+\,J\,--PAS (all tracers; gray).}}
  \label{lcdm_improve}
\end{figure*}

\begin{figure*}[htb!]
  \centering
  \includegraphics[width=0.93\textwidth]{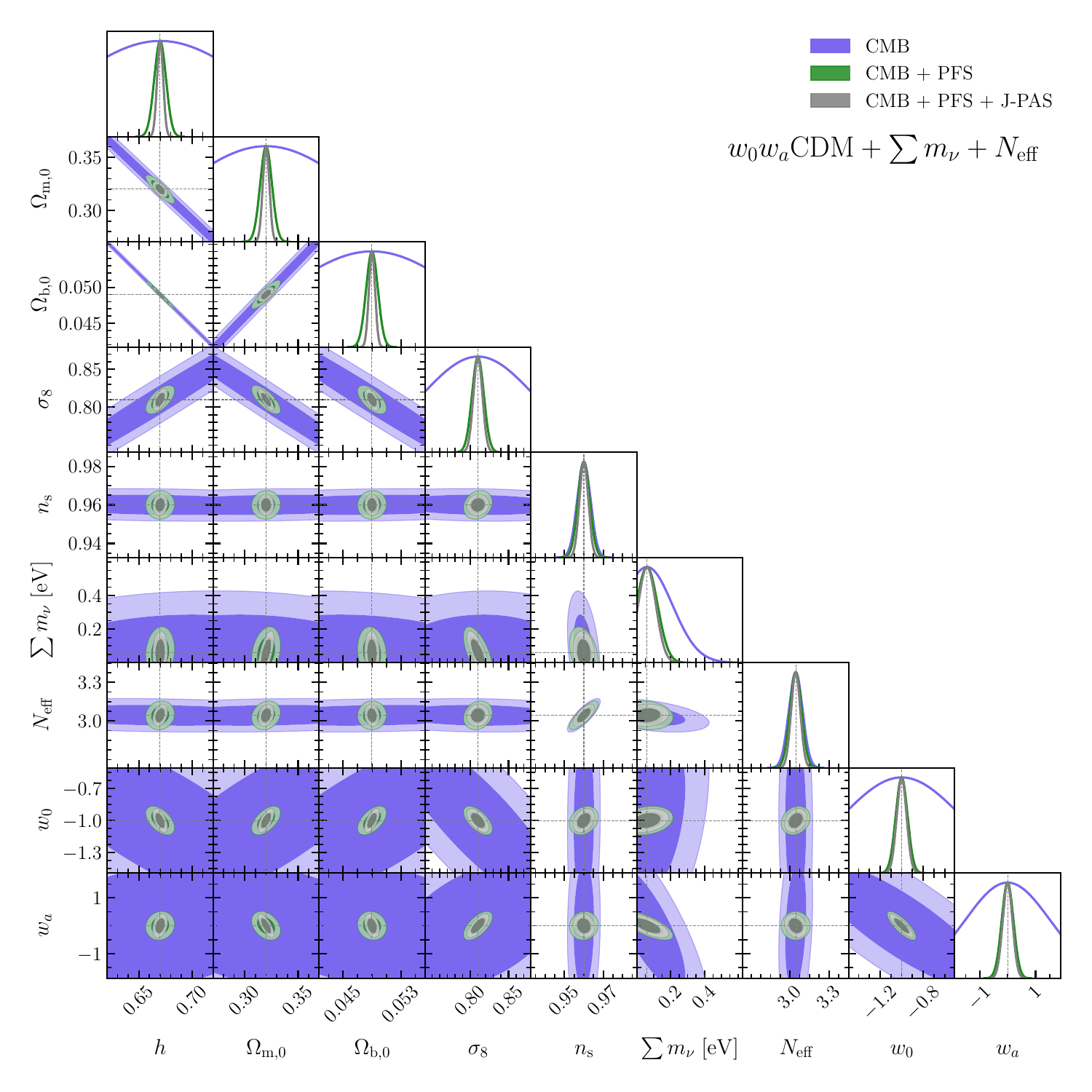}
  \caption{\fx{Same as Figure~\ref{lcdm_improve}, but for the
    $w_0w_a$CDM$+\sum m_\nu+N_{\rm eff}$ cosmology.}}
  \label{w0wacdm_improve}
\end{figure*}

The combination of different galaxy tracers in J-PAS has been shown in Section~\ref{subsec:JPAS} to significantly improve cosmological constraints. Here, we explore the benefit of combining J-PAS ELGs with PFS ELGs, which provide complementary redshift coverage. While J-PAS targets lower redshifts ($z \lesssim 1.2$), PFS maps the Universe up to $z \sim 2.4$ with higher-density ELGs. Their combination spans a broader redshift range ($0.3 < z < 2.2$), allowing us to better leverage redshift evolution and break parameter degeneracies.

Figure~\ref{lcdm_elg_z} illustrates the impact of redshift complementarity in the $\Lambda$CDM$+\sum m_\nu$ model. The left panel shows 68\% confidence contours for ($h$, $\Omega_\mathrm{m,0}$, $\sum m_\nu$) across individual redshift bins from the combination of J-PAS ELG and PFS ELG. The right panel compares the joint constraints from J-PAS ELG, PFS ELG, and their combination. The degeneracy directions change with redshift, and their combination leads to significantly tighter constraints. In particular, the constraint on $\sum m_\nu$ improves by a factor of 2.5 relative to PFS ELG alone and by 1.2 relative to J-PAS ELG alone.

We repeat the analysis in the $w_0w_a$CDM$+\sum m_\nu$ model. Figure~\ref{w0wacdm_elg_z} (left) shows 68\% confidence contours for ($\sum m_\nu$, $w_0$, $w_a$) in individual redshift bins. As redshift increases, the orientation of the contours changes, particularly for the dark energy parameters, demonstrating the redshift evolution of parameter degeneracies. The right panel shows the combined constraints from J-PAS ELG, PFS ELG, and their combination. Compared to PFS ELG alone, the combination improves constraints on ($\sum m_\nu$, $w_0$, $w_a$) by factors of (2.5, 2.2, 2.0), and compared to J-PAS ELG alone by factors of (1.2, 1.2, 1.5).

These results demonstrate that combining J-PAS ELG and PFS ELG leverages the strengths of each survey across different redshift ranges. The synergy between their datasets leads to substantially improved constraints on cosmological parameters, particularly neutrino mass and dark energy evolution.

\subsection{Constraints from Combined Surveys}\label{subsec:joint}

We now assess the constraining power obtained when the galaxy surveys
(J\,--PAS and PFS) are analysed both separately and in combination, and when
these data are further combined with future CMB observations from \fx{SO} and LiteBIRD experiments.
Four cosmological models are considered throughout:
\mbox{$\Lambda$CDM$+\sum m_\nu$},
\mbox{$\Lambda$CDM$+\sum m_\nu+N_{\rm eff}$},
\mbox{$w_0w_a$CDM$+\sum m_\nu$}, and
\mbox{$w_0w_a$CDM$+\sum m_\nu+N_{\rm eff}$}.
Unless otherwise stated, all quoted uncertainties correspond to marginalized
1\,$\sigma$ errors.

Two qualitative trends are evident:
\begin{enumerate}
  \item \textbf{Redshift complementarity for galaxies.}  
        Combining the three J\,--PAS tracers (ELG, LRG, QSO) with the higher-redshift PFS ELG sample
        reduces marginalized errors by 30--50\% relative to either survey
        alone.  The improvement stems from complementary redshift coverage
        ($z\!\lesssim\!1.2$ for J\,--PAS versus $z\!\lesssim\!2.4$ for PFS) and
        from exploiting the multi-tracer technique.  As shown by the
        light-green bars in Figure~\ref{barplot}, the
        J\,--PAS\,+\,PFS combination is already comparable to the full DESI
        forecast (light blue).

  \item \textbf{Galaxy--CMB complementarity.}  
        Adding J\,--PAS\,+\,PFS to CMB measurements from \fx{SO} and LiteBIRD breaks key degeneracies and improves constraints on
        $\sum m_\nu$ and the dark-energy parameters $(w_0,w_a)$ by roughly a
        factor of two (Figures~\ref{lcdm_improve} and~\ref{w0wacdm_improve}).
        The performance of CMB\,+\,J\,--PAS\,+\,PFS approaches that of
        CMB\,+\,DESI in every cosmologcial model considered.
\end{enumerate}

\begin{deluxetable*}{lccccccccc}
\label{tab:results}
  \tablecaption{Marginalized 1\,$\sigma$ uncertainties on cosmological
    parameters for the four cosmological models and the survey combinations
    discussed in the text. Ellipsis (\dots) denote parameters that are not
    varied in the corresponding model.\label{table:numbers}}
  \tabletypesize{\footnotesize}
  \tablehead{\colhead{Survey} & $h$ & $\Omega_{\rm m,0}$ & $\Omega_{\rm b,0}$ &
             $\sigma_8$ & $n_{\rm s}$ & $\sum m_\nu$ [eV] &
             $N_{\rm eff}$ & $w_0$ & $w_a$}
  \startdata
  \multicolumn{10}{c}{$\Lambda$CDM$+\sum m_\nu$}\\
  CMB                  & \fx{0.0068} & \fx{0.0089} & \fx{0.00097} & \fx{0.0101} & \fx{0.0019} & \fx{0.050}  & \dots & \dots & \dots\\
  PFS ELG              & 0.0152 & 0.0131 & 0.00619 & 0.0371 & 0.0464 & 0.553 & \dots & \dots & \dots\\
  J\,--PAS        & 0.0077 & 0.0073 & 0.00329 & 0.0149 & 0.0155 & 0.205  & \dots & \dots & \dots\\
  J\,--PAS\,+\,PFS     & 0.0063 & 0.0048 & 0.00264 & 0.0134 & 0.0118 & 0.184  & \dots & \dots & \dots\\
  DESI           & 0.0057 & 0.0038 & 0.00228 & 0.0138 & 0.0181 & 0.215  & \dots & \dots & \dots\\
  CMB\,+\,J\,--PAS\,+\,PFS & 0.0016 & 0.0021 & 0.00025 & 0.0038 & 0.0015 & 0.017  & \dots & \dots & \dots\\
  CMB\,+\,DESI         & 0.0013 & 0.0018 & 0.00020 & 0.0028 & 0.0015 & 0.013  & \dots & \dots & \dots\\
  \hline
  \multicolumn{10}{c}{$\Lambda$CDM$+\sum m_\nu+N_{\rm eff}$}\\
  CMB                  & \fx{0.0079} & \fx{0.0093} & \fx{0.00110} & \fx{0.0112} & \fx{0.0033} & \fx{0.051}  & \fx{0.048} & \dots & \dots\\
  PFS ELG              & 0.0349 & 0.0157 & 0.00636 & 0.0668 & 0.0652 & 0.673  & 1.769 & \dots & \dots\\
  J\,--PAS                 & 0.0128 & 0.0084    & 0.00340                 & 0.0251       & 0.0161           & 0.297     & 0.585  & \dots & \dots\\
  J\,--PAS\,+\,PFS     & 0.0115 & 0.0060 & 0.00272 & 0.0231 & 0.0125 & 0.269  & 0.532 & \dots & \dots\\
  DESI                 & 0.0080 & 0.0046 & 0.00253 & 0.0175 & 0.0207 & 0.228  & 0.413 & \dots & \dots\\
  CMB\,+\,J\,--PAS\,+\,PFS & 0.0021 & 0.0021 & 0.00028 & \fx{0.0041} & \fx{0.0020} & 0.017  & 0.034 & \dots & \dots\\
  CMB\,+\,DESI         & \fx{0.0017} & 0.0019 & 0.00022 & \fx{0.0031} & 0.0019 & 0.014  & 0.032 & \dots & \dots\\
  \hline
  \multicolumn{10}{c}{$w_0w_a$CDM$+\sum m_\nu$}\\
  CMB                  & \fx{0.0813} & \fx{0.0772} & \fx{0.01190} & \fx{0.0720} & \fx{0.0020} & \fx{0.134}  & \dots & \fx{0.546} & \fx{1.47}\\
  PFS ELG              & 0.0159 & 0.0343 & 0.00842 & 0.0456 & 0.0549 & 0.611  & \dots & 0.275 & 1.07\\
  J\,--PAS               & 0.0101 & 0.0137                  & 0.00398                 & 0.0179       & 0.0185           & 0.216  & \dots   & 0.102     & 0.47    \\
  J\,--PAS\,+\,PFS     & 0.0072 & 0.0114 & 0.00323 & 0.0161 & 0.0147 & 0.196  & \dots & 0.091 & 0.40\\
  DESI                 & 0.0072 & 0.0093 & 0.00262 & 0.0151 & 0.0203 & 0.226  & \dots & 0.046 & 0.27\\
  CMB\,+\,J\,--PAS\,+\,PFS & 0.0027 & \fx{0.0029} & \fx{0.00042} & 0.0059 & 0.0017 & \fx{0.052}  & \dots & 0.044 & 0.17\\
  CMB\,+\,DESI         & 0.0021 & 0.0023 & \fx{0.00033} & 0.0048 & 0.0016 & \fx{0.041}  & \dots & 0.029 & 0.10\\
  \hline
  \multicolumn{10}{c}{$w_0w_a$CDM$+\sum m_\nu+N_{\rm eff}$}\\
  CMB                  & \fx{0.0828} & \fx{0.0787} & \fx{0.01214} & \fx{0.0725} & \fx{0.0035} & \fx{0.150}  & \fx{0.053} & \fx{0.547} & \fx{1.52}\\
  PFS ELG              & 0.0467 & 0.0521 & 0.01123 & 0.1078 & 0.1103 & 0.818  & 2.542 & 0.372 & 1.34\\
  J\,--PAS     & 0.0158 & 0.0145                  & 0.00408                 & 0.0286       & 0.0190           & 0.310     & 0.615     & 0.105  & 0.47   \\
  J\,--PAS\,+\,PFS     & 0.0127 & 0.0128 & 0.00340 & 0.0267 & 0.0160 & 0.284  & 0.556 & 0.094 & 0.41\\
  DESI                 & 0.0116 & 0.0148 & 0.00357 & 0.0282 & 0.0328 & 0.276  & 0.668 & 0.073 & 0.36\\
  CMB\,+\,J\,--PAS\,+\,PFS & 0.0028 & \fx{0.0030} & 0.00042 & 0.0059 & 0.0023 & \fx{0.053}  & \fx{0.035} & \fx{0.045} & \fx{0.17}\\
  CMB\,+\,DESI         & 0.0023 & \fx{0.0026} & 0.00033 & 0.0048 & \fx{0.0020} & \fx{0.044}  & \fx{0.034} & 0.030 & 0.11\\
  \enddata
\end{deluxetable*}

\begin{figure}[htb!]
  \centering
  \includegraphics[width=0.47\textwidth]{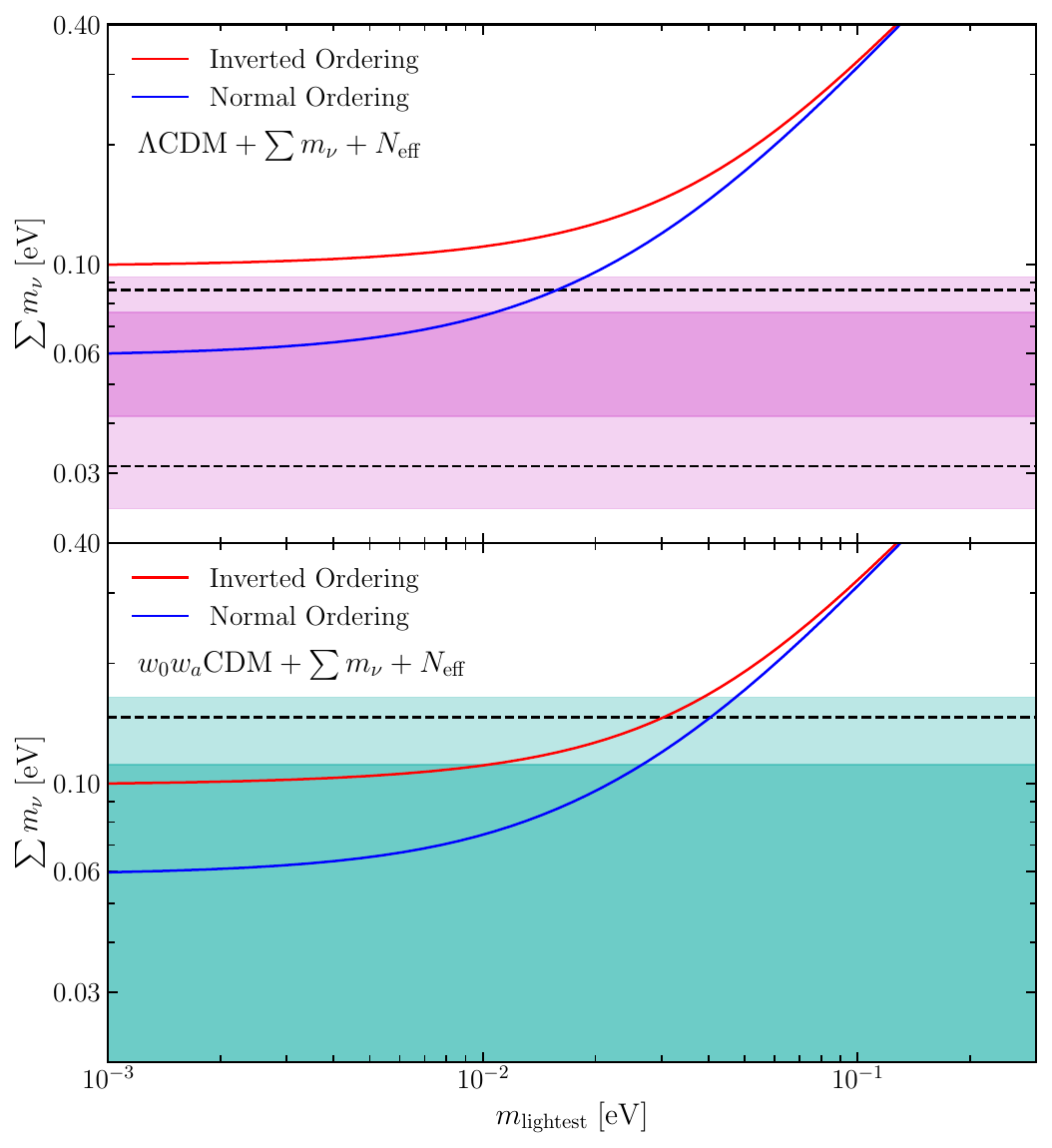}
  \caption{\fx{Forecast constraints on the summed neutrino mass from the data
    combination CMB\,+\,J\,--PAS\,(all tracers)\,+\,PFS\,ELG.
    Solid blue and red curves give the theoretical relations between
    $\sum m_\nu$ and the lightest-neutrino mass $m_{\mathrm{lightest}}$ for the
    normal (NO) and inverted (IO) hierarchies, respectively.
    Shaded bands denote the 68\% (darker) and 95\% (lighter) confidence
    regions for the
    $\Lambda$CDM$+\sum m_\nu+N_{\mathrm{eff}}$ (magenta, upper panel) and
    $w_0w_a$CDM$+\sum m_\nu+N_{\mathrm{eff}}$ (dark-cyan, lower panel)
    forecasts. Dashed lines show the corresponding 95\% limits obtained from
    CMB\,+\,DESI.}}
  \label{band}
\end{figure}

The combination of galaxy surveys with CMB data markedly improves the
constraint on the neutrino-mass sum and therefore offers a potential handle
on the mass hierarchy.  Figure\,\ref{band} compares the forecast limits for
our two cosmological models with the theoretical NO and IO curves. Our work focuses on the parameter constraints achievable by combining CMB and galaxy surveys, whereas the inclusion of supernova data in the joint analysis of CMB and J-PAS survey is discussed in \cite{rodriguesForecasts2025}.

For the $\Lambda$CDM case we obtain
\begin{equation}
  \sigma\!\left(\sum m_\nu\right)=0.017\,\mathrm{eV}
  \quad (\Lambda\mathrm{CDM}+\sum m_\nu+N_{\mathrm{eff}}),
\end{equation}
which disfavors the inverted hierarchy at the
\fx{$2.34\,\sigma$} level if the true mass sum equals the NO minimum.
This implies
$m_{\mathrm{lightest}}<0.019\,\mathrm{eV}$ at 95\% confidence.
For comparison, the combination CMB\,+\,DESI yields
$\sigma(\sum m_\nu)=0.014\,\mathrm{eV}$ and
$m_{\mathrm{lightest}}<0.016\,\mathrm{eV}$, corresponding to a
$2.9\,\sigma$ preference for NO. 

Allowing a time-dependent dark-energy equation of state widens the error bar:
\begin{equation}
  \sigma\!\left(\sum m_\nu\right)=\fx{0.053}\,\mathrm{eV}
  \quad (w_0w_a\mathrm{CDM}+\sum m_\nu+N_{\mathrm{eff}}),
\end{equation}
so the preference for NO drops to \fx{$0.76\,\sigma$}
(\fx{$0.92\,\sigma$} for CMB\,+\,DESI).

\begin{figure}[htb!]
  \centering
  \includegraphics[width=0.47\textwidth]{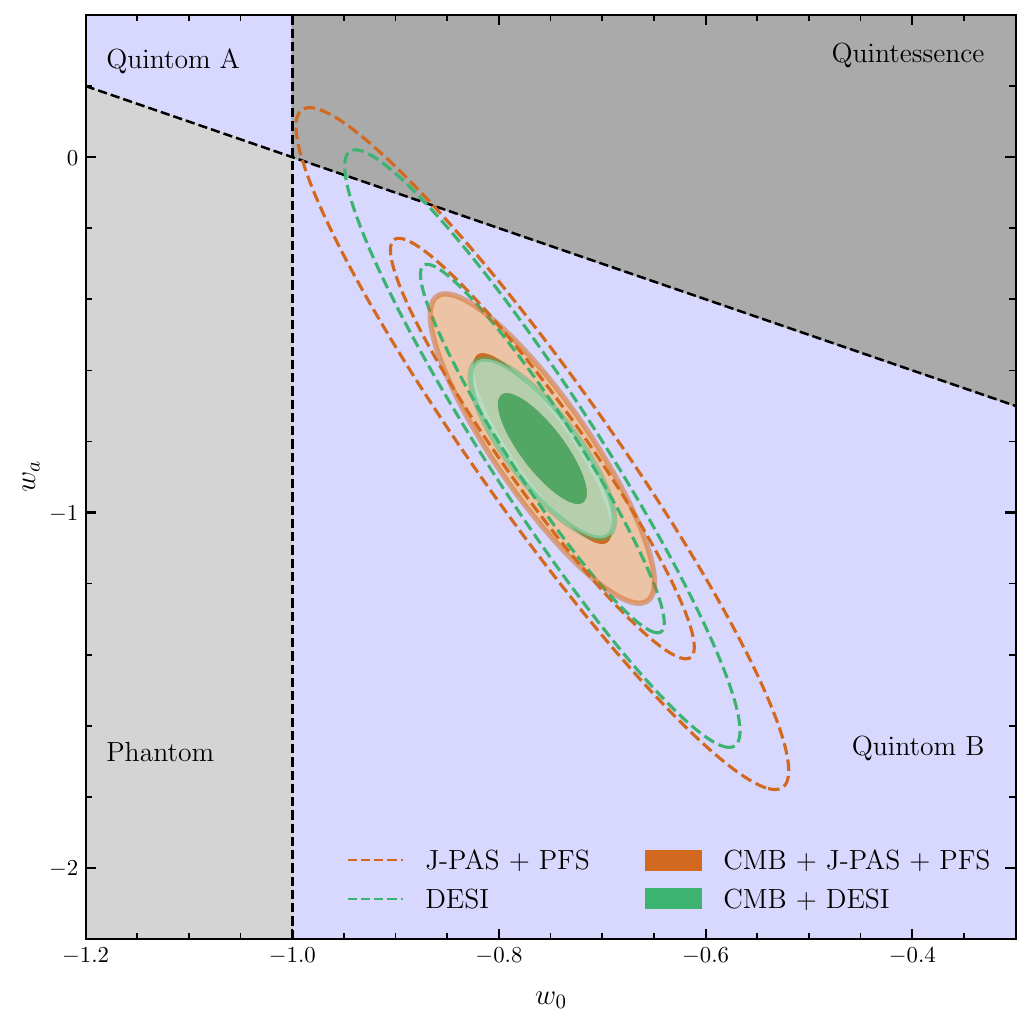}
  \caption{\fx{Forecast 68\% and 95\% confidence contours in the
    $w_0$–$w_a$ plane for the
    $w_0w_a$CDM$+\sum m_\nu+N_{\mathrm{eff}}$ cosmology.
    The parameter space is divided into four quadrants corresponding to
    Quintom\,A ($w>-1$ in the past, $w<-1$ today),
    Quintessence ($w>-1$ at all epochs),
    Phantom ($w<-1$ at all epochs), and
    Quintom\,B ($w<-1$ in the past, $w>-1$ today).
    The intersection of the dashed lines marks the
    $\Lambda$CDM point ($w_0=-1$, $w_a=0$).}}
  \label{de}
\end{figure}

The DESI DR2 analysis, together with Planck CMB and
Type-Ia supernova samples, already prefers a dynamical
dark-energy model at the $2.8$–$4.2\,\sigma$ level,
depending on the supernova compilation employed
\citep{collaborationDESIDR2Results2025}.
To evaluate how upcoming galaxy surveys will sharpen this evidence, we adopt
the DR2 best-fit parameters
$w_0=-0.758$ and $w_a=-0.82$
(Table \uppercase\expandafter{\romannumeral2} of
\citealt{elbersConstraintsNeutrinoPhysics2025})
as our fiducial values and re-run the Fisher forecast within the $\wcdm+\mv+\neff$ model.

For the joint data set
CMB + J\,--PAS (all tracers) + PFS ELG we obtain
\begin{equation}
  \sigma(w_0)=0.044, \qquad
  \sigma(w_a)=0.18.
\end{equation}
The corresponding 68\% and 95\% confidence contours are displayed in
Figure~\ref{de}.
To quantify the tension with $\Lambda$CDM we compute
\begin{equation}
  \Delta\chi^2 =
    (\boldsymbol{\mu}-\boldsymbol{\mu}_{\Lambda{\rm CDM}})^{\!\top}
    \mathrm{Cov}^{-1}
    (\boldsymbol{\mu}-\boldsymbol{\mu}_{\Lambda{\rm CDM}}),
\end{equation}
with $\boldsymbol{\mu}=(-0.758,\,-0.82)$ and
$\boldsymbol{\mu}_{\Lambda{\rm CDM}}=(-1,0)$.
$\mathrm{Cov}$ is the marginalized parameter covariance matrix for ($w_0,w_a$).
Using the forecast covariance matrix we find
\fx{$\Delta\chi^2 = 29.9$}, equivalent to a $5.1\,\sigma$
rejection of $\Lambda$CDM.

For comparison, the combination CMB + DESI (five-year, all tracers) yields
\begin{equation}
  \fx{\sigma(w_0)=0.028,} \qquad
  \sigma(w_a)=0.10,
\end{equation}
tightening the preference for dynamical dark energy to $8.4\,\sigma$.
These results highlight the power of forthcoming wide-field photometric and
spectroscopic surveys to cross-check the emerging DESI indication of
$w(z)\neq-1$ and to pin down the time evolution of dark energy with high
precision.

\section{Summary and Conclusions}\label{conclusion}

We have evaluated the cosmological constraints that the upcoming J\,--PAS and PFS galaxy surveys will deliver, both separately and in combination, and we have quantified the additional gains that arise when these galaxy data are analysed alongside Stage\,--IV CMB observations from \fx{SO} and LiteBIRD.  
Our forecasts target two principal questions: the precision with which the summed neutrino mass, $\sum m_\nu$, can be measured, and the extent to which departures from a cosmological-constant expansion history can be identified through the Chevallier–Polarski–Linder dark-energy parametrisation, $w(a)=w_0+w_a(1-a)$.

The complementary redshift ranges of J\,--PAS (wide, low-$z$) and PFS (dense, high-$z$) break parameter degeneracies that limit either survey on its own.  
Fisher-matrix calculations show that combining their information reduces the marginalised uncertainties on most parameters by 30–50 per cent relative to an analysis of either data set alone.

When the J\,--PAS+PFS combination is added to next-generation CMB temperature and polarisation measurements, the constraints tighten substantially.  
For the $\Lambda$CDM\,$+\sum m_\nu+N_\mathrm{eff}$ model, the full data set is expected to reach  
$\sigma(\sum m_\nu)=0.017\,\mathrm{eV}$,
which would exclude the inverted neutrino mass hierarchy at the \fx{2.34\,$\sigma$} level if the true mass sum equals the normal-ordering minimum.  
Allowing a time-dependent dark-energy equation of state increases the uncertainty on
neutrino mass, yet the joint data set still improves on a CMB-only analysis by more than a factor of two.

Using the DESI DR2 best-fit values, $(w_0,w_a)=(-0.758,-0.82)$, as a fiducial benchmark, the combination of CMB, J\,--PAS, and PFS yields projected uncertainties  
$\sigma(w_0)=0.044$ and $\sigma(w_a)=0.18$,
equivalent to a 5.1\,$\sigma$ preference for dynamical dark energy over $\Lambda$.  
If the final DESI survey is used instead of J\,--PAS and PFS, the significance would increase to about 8.4\,$\sigma$, underscoring the decisive role that future large-scale-structure data will play in testing the physics of cosmic acceleration.

In summary, the forthcoming J\,--PAS and PFS surveys, especially when combined with high-precision CMB measurements, are poised to deliver competitive constraints on neutrino properties and to provide an independent, high-significance probe of dynamical dark energy and some other physics beyond the standard $\Lambda$CDM cosmology.
Their joint performance rivals that anticipated from DESI, ensuring that the next decade of observational cosmology will feature multiple, complementary pathways to precision measurements of fundamental physics.

\begin{acknowledgments}
This work is supported by National Key R\&D Program of China No.(2023YFA1607800, 2023YFA1607803, 2022YFF0503404), NSFC Grants (12273048, 12422301), the CAS Project for Young Scientists in Basic Research (No. YSBR-092), and the Youth Innovation Promotion Association CAS. \fx{AJC acknowledges funding by the European Union - NextGenerationEU and by the Regional Government of Andalucia (AST22\_00001). GR is supported by the Coordenação de Aperfeiçoamento de Pessoal de Nível Superior (CAPES).  FBMS is supported by Conselho Nacional de Desenvolvimento Científico e Tecnológico (CNPq) grant No. 151554/2024-2. JSA is supported by CNPq grant No. 307683/2022-2 and Fundação de Amparo à Pesquisa do Estado do Rio de Janeiro (FAPERJ) grant No. 299312 (2023). JGR acknowledges financial support from the Fundação de Amparo à Pesquisa do Estado do Rio de Janeiro (FAPERJ) grant No. E-26/200.513/2025.}

\fx{Funding for the J-PAS Project has been provided by the Governments of Spain and Arag\'on through the Fondo de Inversiones de Teruel; the Aragonese Government through the Research Groups E96, E103, E16\_17R, E16\_20R, and E16\_23R; the Spanish Ministry of Science and Innovation (MCIN/AEI/10.13039/501100011033 y FEDER, Una manera de hacer Europa) with grants PID2021-124918NB-C41, PID2021-124918NB-C42, PID2021-124918NA-C43, and PID2021-124918NB-C44; the Spanish Ministry of Science, Innovation and Universities (MCIU/AEI/FEDER, UE) with grants PGC2018-097585-B-C21 and PGC2018-097585-B-C22; the Spanish Ministry of Economy and Competitiveness (MINECO) under AYA2015-66211-C2-1-P, AYA2015-66211-C2-2, and AYA2012-30789; and European FEDER funding (FCDD10-4E-867, FCDD13-4E-2685). The Brazilian agencies FINEP, FAPESP, FAPERJ and the National Observatory of Brazil have also contributed to this project. Additional funding was provided by the Tartu Observatory and by the J-PAS Chinese Astronomical Consortium.}
\end{acknowledgments}

\bibliography{draft}{}
%\bibliography{reference}{}
\bibliographystyle{aasjournal}

\end{document}